\documentclass[12pt,twoside]{article}
\usepackage[mathscr]{eucal}
\usepackage{amsmath,amsfonts,amssymb,amsthm}
\usepackage{hyperref}
\usepackage[matrix,arrow]{xy}
\usepackage{times}
\usepackage{graphicx}
\usepackage{epstopdf}
\usepackage{multibox}
\usepackage{dcolumn}

\def\d_Vphi{\text{d}_V\hspace{-0.06em}\phi}
\def\d_Vphibar{\text{d}_V\hspace{-0.06em}\bar\phi}
\def\d_Vxi{\text{d}_V\hspace{-0.06em}\xi}

\def\ndelta{\delta\hspace{-0.50em}\slash\hspace{-0.05em} }
\def\nG{G\hspace{-1.3em}-\hspace{+0.05em} }

\def\be{\begin{eqnarray}}
\def\ee{\end{eqnarray}}
\def\beann{\begin{eqnarray*}}
\def\eeann{\end{eqnarray*}}
\def\beq{\begin{equation}}
\def\eeq{\end{equation}}
\def\ba{\begin{array}}
\def\ea{\end{array}}
\def\ben{\begin{enumerate}}
\def\een{\end{enumerate}}
\def\bea{\begin{eqnarray}}
\def\eea{\end{eqnarray}}

\def\5{\bar }
\def\6{\partial }
\def\7{\hat }
\def\4{\tilde }
\advance\textheight\topskip
\renewcommand{\tilde}{\widetilde}
\renewcommand{\hat}{\widehat}

\newcommand{\bref}[1]{\textbf{\ref{#1}}}

\newcommand{\dd}{\partial}
\renewcommand{\d}{\partial}

\newcommand{\binner}[2]{%
  {\langle}\kern-4.15pt{\langle}#1{,}\,#2{\rangle}\kern-4.15pt{\rangle}}

\newcommand{\half}{\frac{1}{2}}

\newcommand{\ffrac}[2]{\raisebox{.5pt}%
  {\footnotesize$\displaystyle\frac{#1}{#2}$}\kern1pt}

\newcommand{\dover}[2]{\ffrac{\dd #1}{\dd #2}}

\newcommand{\ddl}[2]{\ffrac{\dd #1}{\dd #2}}

\newcommand{\vddl}[2]{{\ffrac{\delta #1}{\delta #2}}}

\def\cD{\mathcal{D}}
\def\cE{\mathcal{E}}
\def\cF{\mathcal{F}}

\def\cH{\mathcal{H}}

\def\cL{\mathcal{L}}

\def\cO{\mathcal{O}}
\def\cP{\mathcal{P}}
\def\cQ{\mathcal{Q}}
\def\cR{\mathcal{R}}

\def\cT{\mathcal{T}}

\def\cW{\mathcal{W}}

\newcommand{\lc}{\boldsymbol{\varepsilon}}
\numberwithin{equation}{section} \makeatletter
\@addtoreset{equation}{section}

\begin{document}

\def\mytitle{Manifest spin 2 duality with electric and magnetic
    sources}

\pagestyle{myheadings} \markboth{\textsc{\small Barnich, Troessaert}}{%
  \textsc{\small Spin 2 duality with sources}} \addtolength{\headsep}{4pt}

\begin{flushright}\small
ULB-TH/08-29
\end{flushright}

\begin{centering}

  \vspace{1cm}

  \textbf{\Large{\mytitle}}



  \vspace{1.5cm}

  {\large Glenn Barnich$^{a}$ and C\'edric Troessaert$^{b}$}

\vspace{.5cm}

\begin{minipage}{.9\textwidth}\small \it \begin{center}
    Physique Th\'eorique et Math\'ematique, Universit\'e Libre de
    Bruxelles\\ and \\ International Solvay Institutes, \\ Campus
    Plaine C.P. 231, B-1050 Bruxelles, Belgium \end{center}
\end{minipage}

\vspace{1cm}

\end{centering}

\vspace{1cm}

\begin{center}
  \begin{minipage}{.9\textwidth}
    \textsc{Abstract}.  We extend the formulation of spin $2$ fields
    on Minkowski space which makes the action manifestly invariant
    under duality rotations to the case of interactions with external
    electric and magnetic sources by adding suitable potentials for
    the longitudinal and trace parts. In this framework, the string
    singularity of the linearized Taub-NUT solution is resolved into a
    Coulomb-like solution. Suitable surface charges to measure
    energy-momentum and angular momentum of both electric and magnetic
    type are constructed.
  \end{minipage}
\end{center}

\vfill

\noindent
\mbox{}
\raisebox{-3\baselineskip}{%
  \parbox{\textwidth}{\mbox{}\hrulefill\\[-4pt]}}
{\scriptsize$^a$Research Director of the Fund for Scientific
  Research-FNRS (Belgium). Also at 
Laboratoire de Math\'ematiques et Physique Th\'eorique,
  Unit\'e Mixte de Recherche du CNRS, F\'ed\'eration Denis Poisson,
  Universit\'e Fran\c cois Rabelais, Parc de Grandmount, 37200 Tours,
  France. \\ $^{b}$Research Fellow of the Fund for Scientific
  Research-FNRS (Belgium).}

\thispagestyle{empty}
\newpage

\begin{small}
{\addtolength{\parskip}{-1.5pt}
 \tableofcontents}
\end{small}
\newpage

\section{Introduction}
\label{sec:introduction}

Duality rotations for massless spin $2$ fields have recently been
shown to be symmetries of the action in the context of a Hamiltonian
formulation involving suitable potentials that arise when solving the
constraints \cite{Henneaux:2004jw}. This symmetry can be extended to
higher spin fields \cite{Deser:2004xt} and also to the case of spin
$2$ fields propagating on an (A)dS background
\cite{Julia:2005ze,Leigh:2007wf}, but does not survive gravitational
self-interactions \cite{Deser:2005sz}. 

The spin $2$ result generalizes the so-called double potential
formalism for spin $1$ fields \cite{Deser:1976iy,Schwarz:1993vs},
which has been extended so as to include couplings to dynamical dyons
by using Dirac strings \cite{Deser:1997mz,Deser:1997se}. In the
original double potential formalism, Gauss's constraint is solved in
terms of new transverse vector potentials for the electric field so
that electromagnetism is effectively formulated on a reduced phase
space with all gauge invariance eliminated. Alternatively, one may
choose \cite{Barnich:2007uu} to double the gauge redundancy of
standard electromagnetism by using a description with independent
vector and longitudinal potentials for the magnetic and electric
fields and $2$ scalar potentials that appear as Lagrange multipliers
for the electric and magnetic Gauss constraints. In this framework,
the string-singularity of the solution describing a static dyon is
resolved into a Coulomb-like solution. Furthermore, magnetic charge no
longer appears as a topological conservation law but as a surface
charge on a par with electric charge.

The aim of the present work is to apply the same strategy to the spin
$2$ case. Doubling the gauge invariance by keeping all degrees of
freedom of symmetric tensors now leads to a second copy of
linearized lapse and shifts as Lagrange multipliers for the new
magnetic constraints. As a consequence, the string singularity of the
gravitational dyon, the linearized Taub-NUT solution is resolved and
becomes Coulomb-like exactly as the purely electric linearized
Schwarzschild solution. Furthermore, as required by manifest duality,
magnetic mass, momentum and Lorentz charges also appear as surface
integrals. 

Our work thus presents a manifestly duality invariant alternative to
\cite{Bunster:2006rt} where the coupling of spin $2$ fields to
conserved electric and magnetic sources has been achieved in a
manifestly Poincar\'e invariant way through the introduction of Dirac
strings.

Recent and not so recent related work includes for instance
\cite{Curtright:1980yk,Nieto:1999pn,Hull:2001iu,Bekaert:2002dt,Bekaert:2002jn,%
  Ellwanger:2006hy,Nurmagambetov:2006vz,deHaro:2008gp,%
  Bergshoeff:2008vc,Boulanger:2008nd,%
  Argurio:2008zt,Argurio:2008nb,Bakas:2008zg} and references therein.

In section~\bref{sec:preliminaries}, we briefly recall the two
ingredients needed for our formulation: the Hamiltonian description of
spin $2$ fields propagating on Minkowski spacetime and the
decomposition of symmetric tensors into their irreducible components,
giving rise to the reduced phase space description of linearized
gravity.

Our analysis starts in section~\bref{sec:action-symmetries} with a
degree of freedom count that shows that the phase space of duality
invariant spin $2$ fields with doubled gauge invariance can be taken
to consist of $2$ symmetric tensors, $2$ vectors and $2$ scalars in
$3$ dimensions. We then define the metric, extrinsic curvature and
their duals in terms of the phase space variables and propose our
duality invariant action principle with enhanced gauge invariance. We
proceed by identifying the canonically conjugate pairs and discuss the
gauge structure, Hamiltonian and duality generators of the theory.  In
the absence of sources, we then show how the generators for global
Poincar\'e transformations of Pauli-Fierz theory, reviewed in detail
in appendix~\bref{sec:poinc-gener-pauli}, can be extended to the
duality invariant theory.

The coupling to external electric and magnetic sources is discussed in
section~\bref{sec:linearized-taub-nut}. The equations of motion are
first solved in the simplest case of a point-particle dyon sitting at
the origin. They are Coulomb-like without string singularities. By
identifying the Riemann tensor in terms of the canonical variables and
computing it for this case, we show in
appendix~\bref{sec:decomp-line-riem} that this solution indeed
describes the linearized Taub-NUT solution. 

In section~\bref{sec:surface-charges} we discuss the surface charges
of the theory and show that they include electric and magnetic
energy-momentum and angular momentum. Because of the non-locality of
the Poisson structure, we proceed indirectly and show that the
expressions obtained by generalizing the surface charges of
Pauli-Fierz theory in a duality invariant way fulfill the standard
properties.  Finally, we investigate how the surface charges transform
under a global Poincar\'e transformation of the sources.

\section{Preliminaries}
\label{sec:preliminaries}

\subsection{Canonical formulation of Pauli-Fierz theory}
\label{sec:canon-form-pauli}

The Hamiltonian formulation of general relativity linearized around
flat spacetime is
\begin{eqnarray}
  \label{eq:2}
  S_{PF}[h_{mn},\pi^{mn},n_m,n]=\int dt\Big[\int
  d^3x\, \big(\pi^{mn}\dot h_{mn}-n^m\cH_m-n\cH \big)-H_{PF}\Big], 
\end{eqnarray}
with
\begin{multline}
H_{PF}[h_{mn},\pi^{mn}]=\int
d^3x\big(\pi^{mn}\pi_{mn}-\half\pi^2+\frac{1}{4}\d^rh^{mn}\d_rh_{mn}-\\
-\half\d_m h^{mn}\d^r h_{rn}+\half\d^m h\d^n h_{mn}-\frac{1}{4}\d^m
h\d_m h\big),\label{H}
\end{multline}
and
\begin{eqnarray}
\cH_m=-2\d^n\pi_{mn},\quad
\cH_\perp=\Delta h-\d^m\d^n h_{mn}.\label{eq:2a}
\end{eqnarray}
Here, indices are lowered and raised with the flat space metric
$\delta_{mn}$ and its inverse, $h={h^m}_m$, $\pi={\pi^m}_m$ and
$\Delta =\d_m\d^m$ is the Laplacian in flat space. The linearized $4$
metric is reconstructed using $h_{00}=-2n$ and $h_{0i}= n_i$.

\subsection{Decomposition of symmetric rank two tensors}
\label{sec:decomp-symm-rank}

Symmetric rank two tensors $\phi_{mn}$ decompose as
\cite{Arnowitt:1962aa,deser:1967aa}
\begin{eqnarray}
\phi_{mn} & = & \phi^{TT}_{mn} + \phi^T_{mn} + \phi^L_{mn},\\
\phi^L_{mn} & = & \partial_m \psi_n + \partial_n \psi_m,\\
\phi^T_{mn} & = & \half \left( \delta_{mn} \Delta-\partial_m \partial_n
\right) 
\psi^T.\label{eq:dec}
\end{eqnarray}
Here $\phi^{TT}_{mn}$ is the transverse-traceless part, containing two
independent components. The tensor $\phi^T_{mn}$ contains the trace
of the transverse part of $\phi_{mn}$ and only one independent
component. The last three components are the longitudinal part
contained in $\phi^L_{mn}$. In terms of the original tensor
$\phi_{mn}$ the potentials for the longitudinal part and the trace are
given by 
\begin{eqnarray}
\psi_{m} & = & \Delta^{-1} \left( \d^n \phi_{mn}
  - \frac{1}{2} \Delta^{-1} \d_m\d^k\d^l\phi_{kl} \right),\\
\psi^T & = & \Delta^{-1} \left( \phi - 
\Delta^{-1} \d^m\d^n\phi_{mn}\right),
\end{eqnarray}
while the transverse-traceless part is then defined as the remainder, 
\begin{eqnarray}
\phi^{TT}_{mn} & = & \phi_{mn} - \phi^T_{mn} - \phi^L_{mn}.
\end{eqnarray}
This implies 
\begin{align}
  \label{eq:26}
\Delta^2
\phi^{TT}_{mn}&=\Delta^2\phi_{mn}-\Delta\d_m\d^k\phi_{kn}-\Delta\d_n\d^k\phi_{km}
\nonumber\\ & -\half\Delta(\delta_{mn}\Delta-\d_m\d_n)\phi +
\half(\delta_{mn}\Delta+\d_m\d_n)\d^k\d^l\phi_{kl},\\
\int d^3x\, \phi^{mn}\Delta^2 \phi^{TT}_{mn} & = \int d^3x\,
\Big(\Delta\phi^{mn}\Delta\phi_{mn}+2 \d_m\phi^{mn}\Delta \d^k\phi_{kn}
-\half (\Delta\phi)^2\nonumber\\ & +\d_m\d_n \phi^{mn}\Delta \phi+\half
\d_m\d_n\phi^{mn}\d_k\d_l\phi^{kl}\Big).
\end{align}

Alternatively, one can introduce the local operator $\mathcal{P}^{TT}$
\begin{equation}
  \left(\mathcal{P}^{TT} \phi \right)_{mn} =
  \half\big[\epsilon_{mpq} \partial^p (\Delta {\phi^{q}}_n
  -\partial_n \partial_r \phi^{qr}) + \epsilon_{npq} \partial^p (\Delta
  {\phi^{q}}_m  - \partial_m \partial_r 
  \phi^{qr})\big], \label{eq:3}
\end{equation}
which projects out the longitudinal and trace parts and onto a
transverse-traceless tensor,
\begin{eqnarray}
\left(\mathcal{P}^{TT} \phi \right)_{mn}  =  \mathcal{P}^{TT} \left(
  \phi^{TT} \right)_{mn}  =
\left( \mathcal{P}^{TT} \phi \right)^{TT}_{mn}. 
\end{eqnarray}
In addition, 
\begin{eqnarray}
  \label{eq:6}
  \left(\mathcal{P}^{TT} ( \mathcal{P}^{TT}\phi)
  \right)_{mn}=-\Delta^3 \phi^{TT}_{mn}.
\end{eqnarray}
As a consequence, the transverse-traceless tensor $\phi^{TT}_{mn}$ can
be written as $\mathcal{P}^{TT}$ acting on a suitable potential
$\psi^{TT}_{mn}$,
\begin{equation}
\phi^{TT}_{mn} =\left( \mathcal{P}^{TT} \psi^{TT}\right)_{mn},
\quad
\psi^{TT}_{mn} = - \Delta^{-3}\left(\mathcal{P}^{TT} \phi
\right)_{mn}.  
\end{equation}

The operator $\mathcal{P}^{TT}$ is related to the way the Hamiltonian
constraint $\cH=0$ is solved by expressing the metric $h_{mn}$ in
terms of superpotentials in \cite{Henneaux:2004jw}. When acting on a
transverse-traceless tensor, the two terms of \eqref{eq:3} involving
$\partial_r \phi^{qr}$ can be dropped. In this case,
$\mathcal{P}^{TT}$ is related to the generalized curl
\cite{Deser:2004xt,Deser:2005sz},
\begin{eqnarray}
  \label{eq:12}
  \left(\cO\phi\right)_{mn}=\half(\epsilon_{mpq} \partial^p {\phi^{q}}_n + 
\epsilon_{npq} \partial^p  {\phi^{q}}_m),\\
\left(\mathcal{P}^{TT} \phi^{TT} \right)_{mn}=\Delta
\left(\cO\phi^{TT}\right)_{mn}. 
\end{eqnarray}
A second operator that projects out the longitudinal and trace
parts and onto a transverse-traceless tensor is $\mathcal{Q}^{TT}$,
\begin{eqnarray}
  \label{eq:8}
  \left(\mathcal{Q}^{TT} \phi \right)_{mn}  =  
\epsilon_{mpq}\epsilon_{nrs} \partial^p\partial^r \Delta\phi^{qs}
-\half(\delta_{mn}\Delta -\d_m\d_n)(\Delta\phi-\d^p\d^r\phi_{pr}). 
\end{eqnarray}
In this case, 
\begin{eqnarray}
  \label{eq:9}
  \left(\mathcal{Q}^{TT} ( \mathcal{Q}^{TT}\phi)
  \right)_{mn}=\Delta^4 \phi^{TT}_{mn},
\end{eqnarray}
so that the transverse-traceless tensor $\phi^{TT}_{mn}$ can
be written as $\mathcal{Q}^{TT}$ acting on another potential
$\chi^{TT}_{mn}$,
\begin{eqnarray}
  \label{eq:10}
  \phi^{TT}_{mn} =\left( \mathcal{Q}^{TT} \chi^{TT}\right)_{mn},
\quad
  \chi^{TT}_{mn} = \Delta^{-4}\left(\mathcal{Q}^{TT} \phi
\right)_{mn}. 
\end{eqnarray}
In turn this operator is related to the way the constraints $\cH_m=0$
are solved by expressing the momenta $\pi^{mn}$ in terms of
superpotentials in \cite{Henneaux:2004jw}.  When acting on a
transverse-traceless tensor, the last term can again be dropped and
it is related to the square of the generalized curl,
\begin{eqnarray}
  \label{eq:13}
  \left(\mathcal{Q}^{TT} \phi^{TT} \right)_{mn}=\Delta 
\left(\cO(\cO\phi^{TT})\right)_{mn}=-\Delta^2 \phi^{TT}_{mn}.
\end{eqnarray}

The elements of the decomposition are orthogonal under 
integration if boundary terms can be neglected,
\begin{equation}
  \int d^3 x\, \phi^{mn} \varphi_{mn} = \int d^3 x 
  \left( \phi^{TTmn} \varphi^{TT}_{mn} +\phi^{Lmn} \varphi^L_{mn} + 
\phi^{Tmn} \varphi^T_{mn} \right).
\end{equation} 
and the operators $\mathcal{P}^{TT},\mathcal{Q}^{TT},\cO $ are
self-ajoint, e.g., 
\begin{eqnarray}
  \label{eq:7}
  \int d^3 x\, \left( \mathcal{P}^{TT} \phi\right)^{mn}\varphi_{mn}=
\int d^3 x\, \phi^{mn}\left( \mathcal{P}^{TT} \varphi\right)_{mn}.
\end{eqnarray}

\subsection{Reduced phase space for linearized gravity}
\label{sec:pauli-fierz-terms}

For completeness, let us briefly recall \cite{Arnowitt:1962aa} the
reduced phase space associated with the Pauli-Fierz action. Because of
the orthogonality of the decomposition, the canonically conjugate
pairs can be directly read off from the kinetic term and are given by
\begin{eqnarray}
  \label{eq:5}
\big (h^{TT}_{mn}( x),\,\pi^{kl}_{TT}(\vec
  y)\big),\quad \big(h^L_{mn}(\vec
  x),\, \pi^{kl}_L( y)\big),\quad 
\big(h^{T}_{mn}( x),\,\pi_{T}^{kl}( y)\big).
\end{eqnarray}
The first class constraints $\cH_m=0=\cH$ are equivalent to
$\pi^{kl}_L=0=h^T_{mn}$. They can be gauge fixed through the
conditions $h^L_{mn}=0=\pi_{T}^{kl}$. The reduced theory only depends
on $2$ degrees of freedom (per spacetime point), the
transverse-traceless components $(h^{TT}_{mn}(\vec
x),\,\pi^{kl}_{TT}( y))$ and the reduced Hamiltonian simplifies
to 
\begin{eqnarray}
  \label{eq:14}
  H^R=\int d^3x\, \Big(
  \pi^{mn}_{TT}\pi_{mn}^{TT}+\frac{1}{4}\d_rh^{TT}_{mn}\d^rh_{TT}^{mn}\Big). 
\end{eqnarray}

\section{Action and symmetries}
\label{sec:action-symmetries}

\subsection{Degree of freedom count}
\label{sec:degree-freedom-count}

In order to be able to couple to sources of both electric and magnetic
type in a duality invariant way, we want to keep all components and
double the gauge invariance of the theory. With 2 degrees of freedom,
$\#\, {\rm dof}=2$, and 8 first class constraints, $\#\, {\rm fcc}=8$,
we thus need $10$ canonical pairs, $\#\, {\rm cp}=10$, according to
the degree of freedom count \cite{Henneaux:1990au}
\begin{eqnarray}
  \label{eq:15}
  2*( \#\, {\rm cp})=2*(\#\, {\rm dof})+2*(\#\, {\rm fcc}). 
\end{eqnarray}
This can be done by taking 2 symmetric tensors, 2 vectors and 2
scalars as fundamental canonical variables, 
\begin{eqnarray}
  \label{eq:18}
  z^A=(H^a_{mn},A^a_m,C^a).
\end{eqnarray} 

\subsection{Change of variables and duality rotations}
\label{sec:change-vari-dula}

For $a=1,2$, consider $h^{a}_{mn}=(h_{mn},h^D_{mn})$ and
$\pi^{mn}_a=(\pi^{mn}_D,\pi^{mn})$ and the definitions
\begin{align}
  \label{eq:16}
  h^a_{mn} &=  \epsilon_{mpq}\d^p {H^{aq}}_{n}+
 \epsilon_{npq}\d^p
 {H^{aq}}_{m}+\d_m A^a_n +\d_n A^a_m +\half
(\delta_{mn}\Delta-\d_m\d_n)C^a\nonumber\\
&=2\Delta^{-1}\left(\cP^{TT} H^a\right)_{mn}+ 
\partial_m \big(\Delta^{-1}\epsilon_{npq} \partial^p  \partial_r
H^{aqr}+A^a_n\big) \nonumber\\ &\hspace*{2cm} + \partial_n\big(
\Delta^{-1} \epsilon_{mpq}    \partial^p \partial_r H^{aqr} +A^a_m\big)
+ \half (\delta_{mn}\Delta-\d_m\d_n)C^a,
\\
\pi^{a}_{mn} &=
\epsilon_{mpq}\epsilon_{nrs} \partial^p\partial^r 
H^{aqs} -\d_m \d^r H^a_{rn} - \d_n \d^r H^a_{rm} -(\delta_{mn}
\Delta-\d_m\d_n) H^a +\delta_{mn} \d^k\d^l H_{kl}^a\nonumber\\ &=
\Delta^{-1}\left(\cQ^{TT} H^a\right)_{mn}
-\d_m \d^r H^a_{rn} - \d_n \d^r H^a_{rm}  
-\half(\delta_{mn}\Delta -\d_m\d_n) H^a\nonumber\\ &\hspace*{2cm}
+\half \Delta^{-1}(\delta_{mn}\Delta +
\d_m\d_n )\d^p\d^rH^a_{pr}
\nonumber\\
 &= -\Delta H^a_{mn}.\label{eq:16b}
\end{align}
The relations for $h_{mn}[H^1,A^1,C^1]$ and $\pi^{mn}[H^2]$ are the
local change of coordinates from the standard canonical variables of
linearized gravity to the new variables. They are invertible and, as
usual, the inverse is not local. The relations for
$h^{2}_{mn}=h^D_{mn}$, $\pi^{mn}_1=\pi^{mn}_D$ serve to denote
convenient combinations of the new variables in terms of which
expressions below will simplify. As indicated by the notation, the
infinitesimal duality rotations among the fundamental variables are
\begin{equation}
  \label{eq:20}
  \delta_D H^a_{mn}=\epsilon^{ab} H_{bmn},\ 
\delta_D A^a_{m}=\epsilon^{ab} A_{bm},\ \delta_D C^a=\epsilon^{ab}
C_{b}.
\end{equation}
Here, $\epsilon_{ab}$ is skew-symmetric with $\epsilon_{12}=1$ and
indices are lowered and raised with $\delta_{ab}$ and its
inverse. Duality invariance will be manifest if all the internal
indices $a$ are contracted with the invariant tensors
$\delta_{ab},\epsilon_{ab}$. Since $h^a_{mn},\pi^{mn}_a$ are
linear combinations of the fundamental variables, they are rotated in
exactly the same way. We can thus consider $h^2_{mn}=h^D_{mn}$,
$\pi^{mn}_1=\pi^{mn}_D$ as the dual spatial metric and the dual
extrinsic curvature in the linearized theory.

\subsection{Action principle and locality}
\label{sec:action-principle}

The duality invariant local action principe that we propose is of the
form
\begin{eqnarray}
  \label{eq:17}
  S_G[z^A,u^\alpha]=\int d^4x\, (a_A[z]\dot z^A  - u^\alpha
  \gamma_\alpha[z])  -\int dt\, H[z],  
\end{eqnarray}
where $u^\alpha$ denote the 8 Lagrange multipliers and $\gamma_\alpha$
the constraints. 

Let us stress here that we use the assumption that the flat space
Laplacian $\Delta$ is invertible in order to show equivalence with the
usual Hamiltonian or covariant formulation of Pauli-Fierz theory and
also to disentangle the canonical structure. The action principle
\eqref{eq:17} itself and the associated equations of motion will be
local both in space and in time independently of this assumption. The
theory itself is not local as a Hamiltonian gauge theory (see
e.g.~\cite{Henneaux:1992ig}, chapter 12) because the Poisson brackets
among the canonical variables will not be local.

\subsection{Canonical structure}
\label{sec:canonical-structure-1}

The explicit expression for the kinetic term is
\begin{multline}
  a_A\dot z^A =
 \epsilon_{ab} H^{amn}\Big(
  \big(\cP^{TT}\dot{H}^b\big)_{mn}+  \d_m\Delta \dot A^b_n
+\d_n\Delta \dot A^b_m+\\+
\half  (\delta^{mn} \Delta-\partial^m \partial^n) \Delta \dot
C^b\Big).
\label{eq:19a}
\end{multline}
The canonically conjugate pairs are identified by writing the
integrated kinetic term as
\begin{multline}
  \int d^4x\, a_A\dot z^A= \int d^4 x \, \Big( -2\Delta\left( \cO
    H^{2}_{TT}\right)^{mn} \dot
  H^{1TT}_{mn}+2\Delta \d_m H^{2mn}_L\dot A^1_n
\\-2\Delta \d_m \dot H^{1 mn}_L A^2_n
 -\half\Delta(\Delta H^{2}_T-\d_{p}\d_q H^{2pq}_{T})\dot
  C^1+\half\Delta (\Delta H^{1}_T-\d_{p}\d_q H^{1pq}_{T})\dot C^2 \Big).
 \end{multline}
This means that the usual canonical pairs of linearized gravity
can be choosen in terms of the new variables as
\begin{eqnarray}
  \label{eq:22}
  \Big( H^{1 TT}_{mn}( x),\,-2
\Delta\left(\cO H^{2}_{TT}\right)^{kl}(
  y)\Big),\,
\Big(C^1( x),\,-\half\Delta(\Delta
 H^{2}_T-\d_{p}\d_q H^{2pq}_{T})
  \,  (y)\Big)\,,
\nonumber \\
\Big(A^1_m ( x),\,
2 \Delta \d_r H^{2rn}_L ( y)\Big),
\end{eqnarray}
The $4$ additional canonical pairs are 
\begin{equation}
  \label{eq:23}
 \Big(A^{2}_m(x),\,
 -2 \Delta \d_r H^{1rn}_L  ( y)\Big),
 \Big(  C^2( x),
  \,\half \Delta(\Delta H^1_T-\d_{p}\d_q H^{1pq}_{T}) ( y)\Big).
\end{equation}
In particular, it follows that
\begin{eqnarray}
  \label{eq:25}
  \{h^a_{mn}(x),\pi^{bkl}(y)\}=\epsilon^{ab}\half(\delta^k_m\delta^l_n+
\delta^k_m\delta^l_n)\delta^{(3)}(x,y).
\end{eqnarray}

\subsection{Gauge structure}
\label{sec:gauge-structure}

The constraints $\gamma_\alpha\equiv( \cH_{am},\cH_{a\perp})$ are choosen as
\begin{eqnarray}
  \label{eq:24}
 \cH_{am}&=&-2\epsilon_{ab}\d^n
 \pi^b_{mn}=2\epsilon_{ab}\Delta\d^nH^b_{mn},\\
\cH_{a\perp}&=&\Delta h_a-\d_m\d_n h_a^{mn}= \Delta^2 C_a.\label{eq:24a}
\end{eqnarray}
They are first class and abelian
\begin{eqnarray}
  \label{eq:12c}
  \{\gamma_\alpha,\gamma_\beta\}=0.
\end{eqnarray}
The constraints $\cH_{1m},\cH_{1\perp}$ are those of the standard
Hamiltonian formulation of Pauli-Fierz theory expressed in terms of
the new variables. The new constraints $\gamma_\Delta^N=0$ are
\begin{equation}
\cH_{2m}=0=\cH_{2\perp}\label{eq:139}.
\end{equation}
They are equivalent to $\d^rH^1_{rm}=0=C^2$
and are gauge fixed through the conditions $A^2_m=0=H^{1T}_{mn}$.
This does not affect $\pi^{2kl}$, while $h^1_{mn}$ is changed by a
gauge transformation. The partially gauge fixed theory corresponds to
the usual Pauli-Fierz theory in Hamiltonian form as described in
section \bref{sec:canon-form-pauli}.

More precisely, the observables of a Hamiltonian field theory with
constraints are defined as equivalence classes of functionals that
have weakly vanishing Dirac brackets with the constraints and where
two functionals are considered as equivalent if they agree on the
surface defined by the constraints (see e.g.~\cite{Henneaux:1992ig}).
The new constraints together with the gauge fixing conditions form
second class constraints. The Dirac bracket algebra of observables of
this (partially) gauge fixed formulation is isomorphic to the Poisson
bracket algebra of observables of the extended formulation on the one
hand, and to the Poisson bracket algebra of observables of Pauli-Fierz
theory on the other hand.

In the same way, the original constraints $\cH_{1m}=0=\cH_{1\perp}$
are equivalent to $\d^rH^2_{rm}=0=C^{1}$ and are gauge fixed through
$A^2_m=0=H^{2T}_{mn}$, leading to the completely reduced theory in
terms of the $2$ transverse-traceless physical degrees of freedom.

If $\varepsilon^\alpha=(\xi^{am},\xi^{a\perp})$
collectively denote the gauge parameters, the gauge symmetries are
canonically generated by the smeared constraints,
\begin{eqnarray}
  \label{eq:27}
 \delta_\varepsilon z^A=\{z^A,\Gamma[\varepsilon]\},\quad
  \Gamma[\varepsilon]=\int d^3x\, \gamma_\alpha\epsilon^\alpha,
\end{eqnarray}
so that 
\begin{equation}
  \label{eq:30}
\delta_\varepsilon
H^a_{mn}=-\Delta^{-1}\epsilon^{ab}(\delta_{mn}\Delta-\d_m\d_n)\xi^\perp_b,\quad 
\delta_\varepsilon A^a_m= \xi^a_m,\quad \delta_\xi C^a=0,
\end{equation}
which implies in particular 
\begin{equation}
\label{eq:37}
\delta_\varepsilon h^{a}_{mn}=\partial_m\xi^a_n+\partial_n\xi^a_m,\qquad 
  \delta_\varepsilon
  \pi^{a}_{mn}=\epsilon^{ab}(\delta^{mn}\Delta-\d^m\d^n)\xi^\perp_b. 
\end{equation}

Note that a way to get local gauge transformations for the fundamental
variables is to multiply the constraints by $\Delta$, which is allowed
when the flat space Laplacian is invertible. This amounts to
introducing suitable potentials for the gauge parameters and Lagrange
multipliers.

\subsection{Duality generator}
\label{sec:duality-generator}

The canonical generator for the infinitesimal duality rotations
\eqref{eq:20} is 
\begin{align}
  \label{eq:1}
  D&=\int d^3x\, \Big( -(\cP^{TT} H^a)_{mn} 
H_a^{mn}+2\Delta \d_r H^{rm}_{a} A^a_{m}\nonumber\\ & \hspace{2cm} -
\half \Delta(\Delta H^a-\d^m\d^n H_{mn}^a) C_a\Big)\nonumber \\ 
&\approx -\int d^3x\,\cP^{TT}(H^a)_{mn} 
H_a^{mn}.
\end{align}
The duality generator is only weakly gauge invariant,
\begin{equation}
  \label{eq:11}
  \{\cH_{am},D\}=\epsilon_{ab} \cH^b_m\qquad \{\cH_{a\perp},D\}=
\epsilon_{ab} \cH^b_\perp.
\end{equation}
On the constraint surface, it coincides with the generator found in
\cite{Henneaux:2004jw} up to normalisation, where it has been cast in
the form of a Chern-Simons term.

\subsection{Hamiltonian}
\label{sec:hamiltonian}

In terms of the new variables \eqref{eq:16}-\eqref{eq:16b}, the
Pauli-Fierz Hamiltonian reads
\begin{multline}
  \label{eq:31}
H_{PF}=\int d^3x\, \Big(H^{amn}\Delta^2
H_{amn}^{TT}-2\Delta\d^r H^2_{rn}\d_s H^{2sn}-\\-\d^r\d^s H_{rs}^2\Delta
H^2-\half (\d^r\d^s H^2_{rs})^2+\frac{1}{8} \Delta C^1 \Delta^2 C^1\Big),
\end{multline}
where one can use \eqref{eq:26} to expand the first term as a local
functional of $H^a_{mn}$. 

The local Hamiltonian $H=\int d^3x\, h$ of the manifestly duality
invariant action principle \eqref{eq:17} is
\begin{align}
  \label{eq:4}
  H &=\int d^3x\, \Big(H^{amn}\Delta^2 
H_{amn}^{TT}-2\Delta\d^r H^a_{rn}\d_s H^{sn}_a-\nonumber\\ 
& \hspace{2cm}-\d^r\d^s H_{rs}^a\Delta
H_a-\half \d^r\d^s H^a_{rs}\d_k\d_l H^{kl}_a 
+\frac{1}{8}\Delta C^a \Delta^2 C_a\Big)\nonumber\\
&=\int d^3x\, \Big(\Delta H^{a}_{mn}\Delta H^{mn}_a-\half \Delta
H^a\Delta H_a +\frac{1}{8}\Delta C^a \Delta^2 C_a\Big).
\end{align}
It is equivalent to the Pauli-Fierz Hamiltonian since it reduces to
the latter when the additional constraints $\d^r H^1_{rm}=0=C^2$
hold. Note that the terms proportional to $\d^r H^a_{rm}$ and $C^a$
may be dropped since they vanish on the constraint surface, $H \approx
\int d^3x\, H^{amn}\Delta^2 H_{amn}^{TT}$.

The Hamiltonian is gauge invariant on the constraint surface,
\begin{equation}
  \label{eq:19}
  \{H,\Gamma[\xi]\}=\int d^3x\, \cH^a_m \d^m \xi^\perp_a.
\end{equation}

In order for the action \eqref{eq:17} to be gauge invariant,
it follows from \eqref{eq:19} that the Lagrange multipliers
$u^\alpha$ need to transform as
\begin{equation}
  \label{eq:21}
  \delta_\xi u^{am}=\dot \xi^{am}-\d^m
   \xi^{a\perp}, \qquad \delta_\xi
  u^{a\perp}=\dot \xi^{a\perp}. 
\end{equation}

\subsection{Poincar\'e generators}
\label{sec:poincare-generators-1}

Consider now a symmetry generator of Pauli-Fierz theory. It is defined
by an observable $K[h^1,\pi^2]$ whose representative is weakly
conserved in time,
\begin{equation}
  \label{eq:54}
  \frac{\d}{\d t} K+\{K,H_{PF}\}\approx 0.
\end{equation}
Since the new Hamiltonian differs from the Pauli-Fierz one by terms
proportional to the new constraints $\gamma^N_\Delta=0$ given
explicitly in \eqref{eq:139}, we have
\begin{equation}
H=H_{PF}+\int d^3x\, \gamma^N_\Delta k^\Delta.\label{eq:55}
\end{equation}
Furthermore, since $K$, when expressed in terms of the new variables, does not
depend on $H^2_L,H^1_T$, so that $\{K,\int d^3x\, \gamma^N_\Delta
k^\Delta\}\approx 0$ in the extended theory, it follows that $K$ is
also weakly conserved and thus a symmetry generator of the extended
theory,
\begin{equation}
  \label{eq:56}
  \frac{\d}{\d t} K+\{K,H\}\approx 0.
\end{equation}

Consider then the Poincar\'e generators $Q_G(\omega,a)$ of Pauli-Fierz
theory as described in Appendix~\bref{sec:poinc-gener-pauli}. When
expressed in terms of the new variables, they are representatives for
the Poincar\'e generators of the extended theory. Indeed, we just have
shown that they are symmetry generators, while we have argued in
Section~\bref{sec:gauge-structure} that their Poisson algebra is
isomorphic when restricted to their respective constraint surfaces.

Since symmetry generators form a Lie algebra with respect to the
Poisson bracket, $\{Q_G(\omega,a),D\}$ is a symmetry generator. In
much the same way as for the Hamiltonian, we now want to show that one
can find representatives for the Poincar\'e generators that are
duality invariant, 
\begin{equation}
\{Q^D_G(\omega,a),D\}=0\label{eq:60},
\end{equation}
by adding terms proportional to the new constraints.

The first step in the proof consists in showing that the reduced phase
space generators, i.e., the generators $Q_G(\omega,a)$ for which all
variables except for the physical $H^a_{TT}$ have been set to zero,
are duality invariant. All other contributions to $Q_G(\omega,a)$ are
then shown to be proportional to the constraints of Pauli-Fierz
theory. Both these steps follow from straightforward but slightly
tedious computations. For the generators of rotations and boosts for
instance the computation is more involved because the explicit $x^i$
dependence has to be taken into account when performing integrations
by parts.

In terms of the new variables, the terms proportional to the
constraints are bilinear in $(h^1,A^2)$, $(\pi^2,A^2)$, $(h^1,C^1)$
and $(\pi^2,C^1)$. The duality invariant generators $Q^D_G(\omega,a)$
are then obtained by adding the same terms with the substitution
$h^1\to h^2$, $A^2\to -A^1$, $\pi^2\to -\pi^1$ and $C^1\to C^2$, while
keeping unchanged the terms involving only the physical variables
$H^a_{TT}$.

As a consequence, the duality invariant Poincar\'e transformations of
$h^1,\pi^2$ are unchanged on the extended constraint surface. They are
given by \eqref{eq:49a}-\eqref{eq:49b} where $\xi^\perp=-{\omega^0}_\nu
x^\nu+a^0$ and $\xi^i=-{\omega^i}_\nu x^\nu +a^i$. Because of
\eqref{eq:60}, those of $h^2,-\pi^1$ are obtained, on the
contraint surface, by applying a duality rotation to the right
hand-sides of \eqref{eq:49a}-\eqref{eq:49b}.

An open question that we plan to address elsewhere is the construction
of the canonical generators for global Poincar\'e transformations in
the presence of both types of sources that will be introduced in the
next section.

\section{Coupling to conserved electric and magnetic sources}
\label{sec:linearized-taub-nut}

\subsection{Interacting variational principle}
\label{sec:coupl-cons-electr}

We define
\begin{equation}
  \label{eq:34}
  h_{0m}^a=n^a_m=h_{m0}^a,\qquad h_{00}^a=-2n^a,
\end{equation}
and consider the action  
\begin{align}
S_T[z^A,u^\alpha;T^{a\mu\nu}]=\frac{1}{16\pi
  G}S_G+S^J\label{eq:61},
\end{align}
with $S_G$ given in \eqref{eq:17} and the gauge invariant
interaction term
\begin{equation}
  \label{eq:35}
  S^J=\int d^4x\, \half h^a_{\mu\nu} T^{\mu\nu}_a,\qquad
  \d_\mu T^{\mu\nu}_a=0,
\end{equation}
where $T^{\mu\nu}_a\equiv (T^{\mu\nu},\Theta^{\mu\nu})$ are external,
conserved electric and magnetic energy-momentum tensors.

\subsection{Linearized Taub-NUT solution}
\label{sec:point-part-grav}

We start by considering the sources corresponding to a point-particle
gravitational dyon with electric mass $M$ and magnetic mass $N$ at
rest at the origin of the coordinate system, for which
\begin{equation}
  \label{eq:62}
  T^{\mu\nu}_a(x)=\delta^\mu_0\delta^\nu_0 M_a\delta^{(3)}(x^i), \qquad
  M_a=(M,N). 
\end{equation}
In this case, only the constraints \eqref{eq:24a} are affected by the
interaction and become
\begin{equation}
  \label{eq:63}
  \cH_{a\perp}=-16\pi GM_a\delta^{(3)}(x). 
\end{equation}
They are solved by 
\begin{equation}
  \label{eq:64}
  \Delta C^a= G M^a(\frac{4}{r}), 
\end{equation}
where $r=\sqrt{x^i x_i}$. 
It is then straightforward to check that all equations of motions are
solved by
\begin{gather}
  \label{eq:66}
C^a=GM^a( 2 r),\quad  
n^a=G M^a(-\frac{1}{r}), \quad
A^a_m=n^{am}=H^a_{mn}=0,\nonumber \\
  h^a_{mn}=GM^a(\delta_{mn}+\frac{x_mx_n}{r^3}),\quad
  \pi_a^{mn}=0.
\end{gather}

The usual Schwarzschild form is obtained after a gauge transformation
with parameter $\xi^{am}=GM^a(-\half \frac{x^m}{r})$,
$\xi^{a\perp}=0$. The solution then reads
\begin{gather}
  \label{eq:82}
  C^a=GM^a(2r),\
  n^a=G M^a(-\frac{1}{r}),\
  A^a_m=GM^a(-\half \frac{x_m}{r}),\ n^{am}=H^a_{mn}=0,\nonumber
  \\
  h^a_{mn}=GM^a(\frac{2 x_mx_n}{r^3}),\quad \pi_a^{mn}=0.
\end{gather}

By computing the Riemann tensor in terms of the canonical variables in
Appendix~\bref{sec:decomp-line-riem}, we show that this solution
describes the linearized Taub-NUT solution. It resolves the string
singularity of the linearized Taub-NUT solution in the standard
Pauli-Fierz formulation. In spherical coordinates, the latter can for
instance be described by
\begin{equation}
  h_{rr}=\frac{2GM}{r}=h_{00},\qquad h_{0\varphi}=-2N(1-\cos\theta),
\label{eq:74}
\end{equation}
and all other components vanishing, with a string-singularity along
the negative $z$-axis.

\section{Surface charges}
\label{sec:surface-charges}

\subsection{Regge-Teitelboim revisited}
\label{sec:regge-teit-revis}

Because the theory is not local as a Hamiltonian gauge theory, the
analysis of surface charges cannot directly be performed as in
\cite{Barnich:2001jy,Barnich:2007bf}. We thus revert to the original
Hamiltonian method of \cite{Regge:1974zd,Benguria:1976in} and adapt it
to the present situation of exact solutions, where there is no need to
discuss fall-off conditions.

Let $\cL_H=a_A \dot z^A-h-\gamma_{\alpha} u^\alpha$, with $h$ a first
class Hamiltonian density and $\gamma_\alpha$ first class constraints
and define $\phi^i=(z^A,u^\alpha)$. Even though it is not so for our
theory, let us first run through the arguments in the case where one
has Darboux coordinates for the symplectic structure, i.e., when
$\sigma_{AB}=\dover{a_B}{z^A}-\dover{a_A}{z^B}$ is the constant
symplectic matrix. We furthermore suppose that we are in a source-free
region of spacetime. In this case one can show that
\begin{equation}
  \label{eq:107b}
  \delta_\varepsilon z^A\vddl{\cL_H}{z^A}+\delta_\varepsilon
  u^\alpha\vddl{\cL_H}{u^\alpha}= -\d_0\Big(\gamma_\alpha
  \varepsilon^\alpha\Big)-\d_is^i_\varepsilon,
\end{equation}
where $s^i_\varepsilon=s^i_\varepsilon[z,u]$ vanishes when the
Hamiltonian equations of motion, including constraints, are satisfied,
$s^i_\varepsilon\approx 0$. This identity merely expresses the general
fact that the Noether current $s^\mu_\varepsilon$ associated to a
gauge symmetry can be taken to vanish when the equations of motions
hold (see e.g.~\cite{Henneaux:1992ig}, chapter 3),
$s^\mu_\varepsilon\approx 0$, and that the integrand of the generator
is given by (minus) the constraints contracted with the gauge
parameters in the Hamiltonian formalism,
$s^0_\varepsilon=-\gamma_\alpha\varepsilon^\alpha$. An explicit
expression for $s^i_\varepsilon$ in terms of the structure functions
can for instance be found in Appendix D
of~\cite{Barnich:2007bf}. Using integrations by parts, one can write
the variations of the constraints under a change of the canonical
coordinates $z^A$ as an Euler-Lagrange derivative, up to a total
derivative,
 \begin{eqnarray}
  \label{eq:29b}
\delta_z(\gamma_\alpha\varepsilon^\alpha) =\delta z^A
\vddl{(\gamma_\alpha\varepsilon^\alpha)}{
    z^A}-\d_i k^{i}_{\varepsilon}.
\end{eqnarray}
where $k^{i}_{\varepsilon}=k^{i}_{\varepsilon}[\delta z,z]$ depends
linearily on $\delta z^A$ and its spatial derivatives.  Taking the
time derivative of \eqref{eq:29b} and using a variation $\delta_\phi$
of \eqref{eq:107b} to eliminate
$\d_0\delta_z(\gamma_\alpha\varepsilon^\alpha)$, one finds
\begin{equation}
  \label{eq:118}
  \d_i\Big(\d_0  k^{i}_{\varepsilon}-\delta_\phi
  s^i_\varepsilon\Big)=
  \d_0\Big(\delta z^A\vddl{(\gamma_\alpha\varepsilon^\alpha)}{
    z^A}\Big)+
  \delta_\phi\Big( \delta_\varepsilon z^A\vddl{\cL_H}{z^A}+\delta_\varepsilon
  u^\alpha\vddl{\cL_H}{u^\alpha}\Big). 
\end{equation}
One now takes $\varepsilon^\alpha_s$ to satisfy $\delta_{\varepsilon_s
}z^A_s=0=\delta_{\varepsilon_s} u^\alpha_s$. Note that in the case of
Darboux coordinates, this also implies that
$\vddl{(\gamma_\alpha\varepsilon^\alpha_s)}{ z^A}=0$. If furthermore
$z^A_s,u^\alpha_s$ is a solution of the Hamiltonian equations of
motion, the RHS of \eqref{eq:118} also vanishes. By using a
contracting homotopy with respect to $\delta \phi^i$ and their spatial
derivatives, one deduces that
\begin{equation}
  \label{eq:119}
  \d_0  k^{i}_{\varepsilon_s}[\delta z,z_s]=(\delta_\phi
  s^i_{\varepsilon_s})[z_s]-\d_j k^{[ij]}_{\varepsilon_s},
\end{equation}
where $k^{[ij]}_{\varepsilon_s}=k^{[ij]}_{\varepsilon_s}[\delta\phi,
\phi_s]$ depends linearily on $\delta \phi^i$ and their spatial
derivatives. Finally, when $\delta z^A_s,\delta u^\alpha_s$ satisfy
the linearized Hamiltonian equations of motion, including constraints,
we find from \eqref{eq:29b} and \eqref{eq:119} that
\begin{equation}
  \label{eq:120}
\d_i k^{i}_{\varepsilon_s}[\delta z_s,z_s]=0,\quad  \d_0
k^{i}_{\varepsilon_s}[\delta z_s,z_s]-\d_j t^{[ij]}_{\varepsilon_s}=0.
\end{equation}
At a fixed time $t=x^0$, consider a closed $2$ dimensional surface
$S$, $\d S=0$, for instance a sphere with radius $r$ and define the
surface charge 1-forms by
\begin{equation}
  \label{eq:121}
  \ndelta \cQ_{\varepsilon_s}[\delta z_s,z_s]=\oint_{S} d^2x_i\,
  k^{i}_{\varepsilon_s}[\delta z_s,z_s], 
\end{equation}
where $d^{2}x_i=\half\epsilon_{ijk}dx^j\wedge dx^k$. The first
relation of \eqref{eq:120} implies that the surface charge 1-form only
depends on the homology class of the closed surface $S$, 
\begin{equation}
  \label{eq:40}
  \oint_{S_1} d^{2}x_m\,k^{m}_{\varepsilon_s}[\delta z_s,z_s]=
\oint_{S_2} d^{2}x_m\,k^{m}_{\varepsilon_s}[\delta z_s,z_s].
\end{equation}
Here $S_1-S_2=\d\Sigma$, where $\Sigma$ is a three-dimensional volume
at fixed time $t$ containing no sources. For instance, the surface
charge $1$-form does not depend on $r$.
The second relation of \eqref{eq:120} implies that it is conserved in
time and so does not depend on $t$ either,
\begin{equation}
  \label{eq:41}
  \frac{d}{dt}\ndelta \cQ_{\varepsilon_s}[\delta z_s,z_s]=0.
\end{equation}
The question is then whether these charge 1-forms are integrable, see
e.g.~\cite{Iyer:1994ys,Barnich:2003xg,Barnich:2007bf} for a
discussion.

\subsection{Linear theories}
\label{sec:linear-theories}

In the case of linear theories, the latter problem does not arise and
the whole analysis simplifies. One can replace \eqref{eq:29b} by
\begin{equation}
  \label{eq:29c}
\gamma_\alpha\varepsilon^\alpha =z^A
\vddl{(\gamma_\alpha\varepsilon^\alpha)}{
    z^A}-\d_i k^{i}_{\varepsilon}[z],
\end{equation}
where $\delta/\delta z^A$ are the (spatial) Euler-Lagrange derivatives
and $k^{i}_\varepsilon[z]$ depends linearily both on the phase space
variables $z^A$ and their spatial derivatives and on the gauge
parameters. One then uses \eqref{eq:107b} directly to eliminate
$\d_0(\gamma_\alpha\varepsilon^\alpha)$ from the time derivative of
\eqref{eq:29c}, to get
\begin{equation}
  \label{eq:108}
  \d_i\Big[\d_0 k^i_\varepsilon-s^i_\varepsilon\Big]=\d_0\Big[z^A
  \vddl{(\gamma_\alpha\varepsilon^\alpha)}{
    z^A}\Big]+\delta_\varepsilon
  z^A\vddl{\cL_H}{z^A}+\delta_\varepsilon
  u^\alpha\vddl{\cL_H}{u^\alpha}.
\end{equation}
For gauge parameters $\varepsilon^\alpha_s$ that satisfy
\begin{equation}
\delta_{\varepsilon_s }z^A=0=\delta_{\varepsilon_s} u^\alpha,\label{eq:127}
\end{equation}
one then arrives at
\begin{equation}
  \label{eq:122}
 \d_i k^{i}_{\varepsilon_s}[z]=-\gamma_\alpha\varepsilon^\alpha,\quad  \d_0
k^{i}_{\varepsilon_s}[z]=s^i_{\varepsilon_s}[z,u]-\d_j
k^{[ij]}_{\varepsilon_s}. 
\end{equation}
For a solution $z^a_s,u^\alpha_s$, the surface charges
\begin{equation}
\cQ_{\varepsilon_s}[z_s]=\oint_{S} d^2x_i\,
  k^{i}_{\varepsilon_s}[z_s],\label{eq:123}
\end{equation}
are again independent of $r$ and $t$. 

When this analysis is applied to the Hamiltonian formulation of
Pauli-Fierz theory, one finds the standard expressions 
\begin{equation}
k^{i}_{\varepsilon}[z] =2\xi_m\pi^{mi}-
  \xi^{\perp}(\delta^{mn}\d^i-\delta^{mi}\d^n) h_{mn}+
  h_{mn}(\delta^{mn}\d^i-\delta^{ni}\d^m)\xi^{\perp},\label{eq:128}
\end{equation}
while the only solutions to \eqref{eq:127} are $\xi_{\mu
  s}=-\omega_{[\mu\nu]}x^\nu +a_\mu$, for some constants $a_\mu$,
$\omega_{[\mu\nu]}=-\omega_{[\nu\mu]}$. In this context of
flat space, Greek indices take values from $0$ to $3$ with
$\mu=(\perp,i)$. Indices $\mu$ are lowered and raised with
$\eta_{\mu\nu}=\text{diag} \,(-1,1,1,1)$. 

\subsection{Electric and magnetic energy-momentum and angular momentum surface charges }
\label{sec:electr-magn-energy}

The previous analysis is not directly applicable to our case since we
do not have Darboux coordinates and the Poisson brackets of the
fundamental variables are non-local. In particular, the gauge
transformations \eqref{eq:30} do not allow for non trivial solutions
to $\delta_{\varepsilon_s }z^A=0$. We also have to keep the sources
explicitly throughout the argument, because $\Delta^{-1}$ applied to
localized sources will spread them out throughout space and we need to
check that we are only dropping terms that indeed vanish outside of
the sources.

In the presence of the sources, the constraints
$\gamma_\alpha^J=(\cH_{am}^J,\cH_{a\perp}^J)$ are determined
\begin{equation}
\cH_{am}^J=\cH_{am}-(16\pi G) T_{am}^0,\quad
\cH_{a\perp}^J=\cH_{a\perp} -(16\pi G) T_{a0}^0.\label{eq:38}
\end{equation}
Instead of \eqref{eq:29c}, we can write
\begin{multline}
  \label{eq:105}
  \gamma^J_\alpha\varepsilon^\alpha 
  =(\d^m\xi^{an}+\d^n\xi^{am})\epsilon_{ab} \pi^b_{mn}+
  (\delta^{mn}\Delta-\d^m\d^n)\xi^{a\perp} h_{amn} -\d_i
 \tilde k^{i}_{\varepsilon}[z]\\ -(16\pi G)
  (T_{am}^0\xi^{am}+T_{a0}^0\xi^{a\perp}) ,
\end{multline}
where
\begin{equation}
  \tilde k^{i}_{\varepsilon}[z] =2\xi^a_m\epsilon_{ab}\pi^{bmi}-
  \xi^{a\perp}(\delta^{mn}\d^i-\delta^{mi}\d^n) h_{amn}+
  h_{amn}(\delta^{mn}\d^i-\delta^{ni}\d^m)\xi^{a\perp}.
\end{equation}
Consider now gauge parameters $\epsilon^\alpha_s(x)$ 
satisfying the conditions
\begin{equation}
  \label{eq:125}
  \left\{\begin{array}{c}\d^m\xi^{an}_s+\d^n\xi^{am}_s=0=
      \d_0\xi^{am}_s-\d^m
      \xi^{a\perp}_s,  \\
      (\delta^{mn}\Delta-\d^m\d^n)\xi^{a\perp}_s=0=
 \d_0 \xi^{a\perp}_s, \end{array}\right.
\end{equation}
The general solution to conditions \eqref{eq:125} can be written as
\begin{equation}
  \label{eq:43}
  \xi^{a}_{\mu s}=-\omega^{a}_{[\mu\nu]}x^\nu+a^{a}_\mu,
\end{equation}
for some constants $a^{a}_\mu$,
$\omega^{a}_{[\mu\nu]}=-\omega^{a}_{[\nu\mu]}$.
It follows in particular that the surface charges 
\begin{equation}
\cQ_{\varepsilon_s}[z_s]=\frac{1}{16\pi
  G}\oint_{S} d^2x_i\,
 \tilde k^{i}_{\varepsilon_s}[z_s],\label{eq:104}
\end{equation}
do not depend on the homology class of $S$ outside of sources.

Assuming $\Delta$ invertible, the equations of motion associated to
$\cL_T=\frac{1}{16\pi G}\cL_H+\cL^J$ imply in particular that
\begin{multline}
  \label{eq:94}
  \d_0 h^a_{mn}=\d_m n_n^a+\d_nn_m^a-2\epsilon^{ab}\Delta
  H_{bmn}+\epsilon^{ab}\delta_{mn}\Delta H_b \\ 
  +(16\pi G) \epsilon^{ab}\Big(\Delta^{-1} \left(\cO
    T_b\right)_{mn}+\half \Delta^{-2}\d_m\epsilon_{npq}\d^p\d_k
  T^{kq}_b+\half \Delta^{-2}\d_n\epsilon_{mpq}\d^p\d_k T^{kq}_b\Big),
\end{multline}
\begin{multline}
 \epsilon_{ab} \d_0 \pi^b_{mn} =\left(\cP^{TT} H_a\right)_{mn}+(8\pi
  G) T_{amn}-\\-\half (\delta^{mn}\Delta-\d^m\d^n) (2n_a+\half \Delta C_a).
\end{multline}
By direct computation using the equations of motion, one then finds
\begin{equation}
  \label{eq:105a}
  \d_0 \tilde k^i_{\varepsilon_s}[z_s]=(16\pi G)(\xi^a_{\mu s} T_a^{\mu
    i})-\d_j k^{[ij]}_{\varepsilon_s}[z_s,u_s],
\end{equation}
with 
\begin{multline}
  \label{eq:107}
  k^{[ij]}_{\varepsilon_s}[z,u]=\Big(2n_a^i\d^j\xi^{a\perp}_s+\xi^{a\perp}_s\d^i
  n_a^j+\xi^{ai}_s\d^j(2n_a+\half \Delta C_a)+\xi^{a}_{s
    m}\epsilon^{mpq}\d_p\d^i H^j_{aq}\\+ \omega^{aj} \d^k
  H^i_{ak}+ \omega^{ai}\d^jH_a+2\omega^{ak}\d^i H^j_{ak}+16\pi
  G\epsilon^{ab}\epsilon^{imq}\Delta^{-1} T_{bq}^j\d_m\xi_{as}^\perp
  \\+8\pi G\epsilon^{ab}\epsilon^{mpq}\d_p\Delta^{-2} \d^i
  T^j_{bq}\d_m\xi_{as}^\perp -(i\longleftrightarrow j)\Big)
  \\+\epsilon^{ijk}\Big[\omega^a_k(2n_a+\half \Delta C_a) -
  \xi^{am}_s(\Delta H_{amk}-\d_m\d^rH_{ark})\\-16\pi
  G\epsilon^{ab}\Delta^{-1} \d^r T_{brk}\xi_{as}^\perp+8\pi
  G\epsilon^{ab}(\Delta^{-1}T^m_{bk}+\Delta^{-2}\d^m\d^rT_{brk})\d_m\xi^\perp_{as}
  \Big],
\end{multline}
where $\omega^a_{mn}=\omega^{ak}\epsilon_{kmn}$. The surfaces charges
\eqref{eq:104} are thus also time-independent outside of sources. 

Finally, the surface charges are gauge
invariant,
\begin{gather}
  \tilde k^i_{\varepsilon_s}[\delta_{\eta}z]=\d_j
  r^{[ij]}_{\varepsilon_s,\eta}, \\
r^{[ij]}_{\varepsilon_s,\eta}=\Big(2\xi^{aj}_{s}\d^i\eta^\perp_a+
2\eta_a^j\d^i\xi^{a\perp}_s+\xi^{a\perp}_s\d^j
  \eta_a^i-(i\longleftrightarrow j)\Big)
-2\epsilon^{ijk}\omega^a_k\eta^\perp_a. 
\label{eq:112}
\end{gather}

Defining 
\begin{equation}
Q_{\epsilon_s}[z]=\half\omega^a_{\mu\nu}J^{\mu\nu}_a-a^a_\mu
P^\mu_a,
\label{eq:45}
\end{equation}
we get for the individual generators
\begin{eqnarray}
(16\pi G)  P_a^\perp & = & -\oint_{S^\infty} d^2x_m \, \partial^m \Delta C_a = 
\oint_{S^\infty} d^2x_m \,\left( \partial_nh^{mn}_a - \partial^mh_a\right),\\
(16\pi G)  P_a^n & = & 2\oint_{S^\infty} d^2x_m \, \epsilon_{ab} 
\Delta H^{bnm} =- 2\oint_{S^\infty} d^2x_m \, \epsilon_{ab} \pi^{bmn},\\
(16\pi G)  J_a^{kl} & = & 2\oint_{S^\infty} d^2x_m \, 
\epsilon_{ab} \left( \Delta H^{bmk}x^l-\Delta H^{bml}x^k\right) \\ &&
\quad = -2\oint_{S^\infty} d^2x_m \, \epsilon_{ab} 
\left( \pi^{bmk}x^l-\pi^{bml}x^k\right),\\
(16\pi G)  J^{\perp k}_a &= & \oint_{S^\infty} d^2x_m \, \left( \Delta C^a
    \delta^{mk} - \partial^m \Delta C_a x^k\right)\\ && \quad =
  \oint_{S^\infty} d^2x_m \, \left[ \left(\partial_n h^{mn}_a
      - \partial^m h_a \right) x^k -h^{mk}_a + h_a 
    \delta^{mk}\right].
\end{eqnarray}

The only non-vanishing surface charges of the dyon sitting at the
origin are  
\begin{equation}
 P^\perp_a= M_a.\label{eq:68}
\end{equation}
As expected, they measure the electric and magnetic mass of the dyon.

For later use, we combine $\tilde
k^i_\varepsilon,k^{[ij]}_{\varepsilon}$ into the $n-2$ forms
$k_\varepsilon[z,u]$ through the following expressions in Cartesian
coordinates, 
\begin{gather}
  k_{\varepsilon}=k^{[\mu\nu]}_\varepsilon d^{n-2}x_{\mu\nu},\qquad 
  k^{[0i]}_\varepsilon =\tilde k^i_\varepsilon,\\
  d^{n-k}x_{\mu_1\dots\mu_k}=\frac{1}{k!(4-k)!}
  \epsilon_{\mu_1\dots\mu_k\nu_{k+1}\dots
    \nu_4}dx^{\nu_{k+1}}\dots dx^{\nu_{4}},\label{eq:135} \end{gather}
where $\epsilon_{\alpha\beta\gamma\delta}$ is completely
skew-symmetric with $\epsilon_{0123}=1$ and the wedge product between
the differentials is understood. Equations \eqref{eq:105} and
\eqref{eq:105a} can then be summarized by 
\begin{equation}
  \label{eq:136}
  d k_{\varepsilon_s}\approx -(16\pi G) T_{\varepsilon_s},\quad
T_{\varepsilon_s}=T^{\mu}_{a\nu}\xi^{a\nu}_sd^{3}x_\mu,\quad
dT_{\varepsilon_s}=0,
\end{equation}
where closure of the $n-1$-forms $T_{\varepsilon_s}$ follows from the
conservation of the sources, the symmetry of the energy-momentum
tensor and \eqref{eq:43}. 

{\bf Remark:} In fact we have checked here that the standard
expressions for surface charges in Pauli-Fierz theory, when extended
in a duality invariant way, have all the expected properties. More
interesting would be to develop the theory of surface charges from
scratch in theories where the Poisson brackets among the fundamental
variables are not local to see if the ones we have found exhaust all
possibilities. From the preceding discussion we see that
pseudo-differential operators will play a crucial role for a
discussion of these generalized conservation laws, as they do in the
discussion of ordinary conservation laws for evolution equations of
the Korteweg-de Vries type for instance.

\subsection{Poincar\'e transformations of surface charges}
\label{sec:poinc-transf-pres}

Suppose now that $z^A_s,u^\alpha_s$ solve the equations of motions for
the conserved sources $T^{\mu\nu}_a(x)$. Let $z^{\prime A}_s,u^{\prime
  \alpha}_s$ be the solution associated to new sources $T^{\prime
  \mu\nu}_a(x)$ related to $T^{\mu\nu}_a(x)$ through a (proper)
Poincar\'e transformation, $x^{\prime\mu}={\Lambda^\mu}_\nu
x^\nu+b^\mu$ with $|\Lambda|=1$, \begin{equation}
  \label{eq:130}
  T^{\prime\mu\nu}_a(x^\prime)={\Lambda^\mu}_\alpha{\Lambda^\nu}_\beta
  T^{\alpha\beta}_a(x).  
\end{equation}
For instance, starting from the conserved energy-momentum tensors
\eqref{eq:62} of a dyon sitting at the origin with world-line
$z^\mu=\delta^\mu_0 s$, one can obtain in this way the conserved
energy-momentum tensors of a dyon moving along a straight line,
$z^{\prime\mu}=u^\mu s+ a^\mu$ with $u^\mu,a^\mu$ constant, $u^\mu
u_\mu=-1$ and $s$ the proper time,
\begin{equation}
  \label{eq:113}
  T^{\prime\,\mu\nu}_a(x^\prime)=M_a u^\nu\int d\lambda \delta^{(4)}(x^\prime-
z^\prime(\lambda))  \frac{dz^{\prime\mu}}{d\lambda}=M_a\frac{u^\mu
      u^\nu}{u^0}\delta^{(3)}(x^{\prime i}-z^{\prime i}(x^0)).    
\end{equation}
when ${\Lambda^\mu}_0=u^\mu$.

Assume then that the $\xi^\mu_{a s}(x)$ transform like vectors
\begin{equation}
\xi^{\prime\nu}_{a
  s}(x^\prime)={\Lambda^\mu}_\alpha\xi^{\alpha}_{as}(x)
=-(\Lambda \omega_a\Lambda^{-1}
x^\prime)^\nu+(\Lambda \omega_a\Lambda^{-1}b+\Lambda a_a)^\nu,
\label{eq:137}
\end{equation}
which implies that the $T_{\varepsilon_s}$ are closed Poincar\'e invariant
$n-1$ forms, 
\begin{equation}
  \label{eq:133}
  T^{\prime}_{\varepsilon^\prime_s}(x^\prime,dx^\prime)=
  T_{\varepsilon_s}(x,dx). 
\end{equation}
We can then use the following variant of the tube lemma. Suppose that
at fixed time $x^0$, $T^{0\nu}_a(x)\xi^{a\nu}_s(x)$ has compact
support and that there exists a tube, i.e., a space-time volume $\cW$
connecting the hypersurfaces $\Omega: K=x^0$ and $\Omega^\prime:
K=x^{\prime 0}={\Lambda^0}_\nu x^\nu+ b^0$ with $K$ a constant such
that, $\partial \cW=\Omega^\prime-\Omega+\cT$. If nothing flows out through
$\cT$, $\int_{\cT} T_{\varepsilon_s}=0$, it follows from Stokes'
theorem and \eqref{eq:133} that
\begin{equation}
  \label{eq:138}
  \int_\Omega T_{\varepsilon_s}= \int_{\Omega^\prime} 
T_{\varepsilon_s}=\int_{\Omega^\prime}
  T^\prime_{\varepsilon^\prime_s}. 
\end{equation}
If we now compute the surface charges for a large enough sphere $S$ at
fixed $x^0$ containing both $T^{0\nu}_a(x)\xi^{a\nu}_s(x)$ and
$T^{\prime 0\nu}_a(x)\xi^{\prime a\nu}_s(x)$, it finally follows from
\eqref{eq:105} or \eqref{eq:136} that the surface charges evaluated
for the new solutions $z^{\prime A}$ are obtained from those of the
old solutions $z^A$ through
\begin{equation}
  \label{eq:132} \cQ_{\varepsilon^\prime_s}[z^\prime_s]=\cQ_{
    \varepsilon}[z_s]. 
\end{equation}

\section*{Acknowledgements}
\label{sec:acknowledgements}

\addcontentsline{toc}{section}{Acknowledgments}

The authors thank R.~Argurio, X.~Bekaert, F.~Dehouck, A.~Gomberoff,
and L.~Houart for useful discussions. This work is supported in part
by a ``P{\^o}le d'Attraction Interuniversitaire'' (Belgium), by
IISN-Belgium, convention 4.4505.86, by the Fund for Scientific
Research-FNRS (Belgium), and by the European Commission programme
MRTN-CT-2004-005104, in which the authors are associated to
V.U.~Brussel.

\appendix

\section{Poincar\'e generators for Pauli-Fierz theory}
\label{sec:poinc-gener-pauli}

In this appendix, we assume that the canonical variables vanish
sufficiently fast at the boundary so that integrations by parts can be
used even if the gauge parameters do not vanish at the boundary. 

In the Hamiltonian formulation of general relativity
\cite{Arnowitt:1962aa}, the canonically conjugate variables are the
spatial $3$ metric $g_{ij}$ and the extrinsic curvature
$\pi^{ij}$. The constraints are explicitly given by
\begin{equation}
  \mathcal{H}_\perp  =  
\frac{1}{\sqrt{g}}\left( \pi^{mn}\pi_{mn} - \frac{1}{2}\pi^2\right)-
\sqrt{g} R\,,\qquad 
  \mathcal{H}_i  =  -2 \nabla_j \pi^{j}_{\phantom{j}i}. 
\end{equation}
The associated generators of gauge transformation $H[\xi]=\int d^3x \,
\left( \mathcal{H}_\perp \xi^\perp + \mathcal{H}_i \xi^i\right)$
satisfy the so-called surface deformation algebra
\cite{Teitelboim:1972vw,Hojman:1976vp},
\begin{eqnarray}
\label{Halgebra}
\left\{ H[\xi] , H[\eta]\right\} & = & H[[\xi,\eta]_{SD}],\\
{[\xi,\eta]}^\perp_{SD} & = & \xi^i \partial_i \eta^\perp 
- \eta^i \partial_i \xi^\perp,\\
{[\xi, \eta ]}^i_{SD} & = & g^{ij}\left( \xi^\perp \partial_j
  \eta^\perp - \eta^\perp \partial_j \xi^\perp\right) 
+ \xi^j \partial_j \eta^i - \eta^j \partial_j \xi^i.
\end{eqnarray}
When the parameters $f,g$ of gauge transformations depend on the
canonical variables, \eqref{Halgebra} is replaced by
\cite{Brown:1986nw}
\begin{align}
  \label{eq:32}
  \left\{ H[f] , H[g]\right\} & =  H[k],\\
k& =[f,g]_{SD}+\delta_g f-\delta_f g -m, \\
m^\perp & =\int d^3x^\prime\, \Big[\{f^\perp, g^\perp(x^\prime)\}
\cH_{\perp}(x^\prime) + \{f^\perp, g^j(x^\prime)\}
\cH_{j}(x^\prime)\Big],\\
m^i & =\int d^3x^\prime\, \Big[\{f^i, g^\perp(x^\prime)\}
\cH_{\perp}(x^\prime) + \{f^i, g^j(x^\prime)\}
\cH_{j}(x^\prime)\Big],
\end{align}
where 
\begin{align}
  \label{eq:47}
  \delta_\xi g_{ij} & = \nabla_i \xi_j+\nabla_j\xi_i+2
  D_{ijkl}\pi^{kl}\xi^\perp,\\
D_{ijkl}&=\frac{1}{2\sqrt g}(g_{ik}g_{jl}+g_{jk}g_{il}-g_{ij}g_{kl}),\\
\delta_\xi \pi^{ij} & = -\xi^\perp\sqrt g(R^{ij}-\half
g^{ij}R)+\frac{\xi^\perp}{2\sqrt
  g}g^{ij}(\pi^{kl}\pi_{kl}-\half\pi^2)\nonumber\\ & -2\frac{\xi^\perp}{\sqrt g}
(\pi^{im}\pi_m^j -\half \pi^{ij}\pi)+\sqrt
g(\nabla^j\nabla^i\xi^\perp-g^{ij}\nabla_m \nabla^m \xi^\perp)\nonumber\\ &
+\nabla_m (\pi^{ij} \xi^m)-\nabla_m
\xi^i\pi^{mj}-\nabla_m\xi^j\pi^{mi}. \label{eq:47a}
\end{align}

Let $g_{ij}=\delta_{ij}+h_{ij}$ and consider the canonical change of
variables from $g_{ij},\pi^{kl}$ to $z^A=(h_{ij},\pi^{kl})$. We will
expand in terms of the homogeneity in the new variables and use the
flat metric $\delta_{ij}$ to raise and lower indices in the remainder
of this appendix. Furthermore, Greek indices take values from $0$ to
$3$ with $\mu=(\perp,i)$. Indices are lowered and raised with
$\eta_{\mu\nu}={\rm diag} (-1,1,1,1)$ and its inverse. Let
$\tilde\omega_{\mu\nu}=-\tilde\omega_{\nu\mu}$.

To lowest order, i.e., when $g_{ij}=\delta_{ij} $, the vector fields 
\begin{equation}
\label{vect}
\xi_P(\tilde\omega,\tilde a)^\mu=
-\tilde\omega^\mu_{\phantom{\mu}i}x^i+\tilde a^\mu,
\end{equation}
equipped with the surface deformation bracket form a representation of
the Poincar\'e algebra \cite{Regge:1974zd},
\begin{equation}
\label{vectalgebra}
[\xi_P(\tilde \omega_1,\tilde a_1), \xi_P(\tilde \omega_2,\tilde 
a_2)]^{(0)}_{SD}=
\xi_P([\tilde \omega_1,\tilde\omega_2], \tilde\omega_1 \tilde a_2 -
\tilde\omega_2 \tilde a_1).
\end{equation}

For the gauge generators, we find $H[\xi]=H^{(1)}[\xi] + H^{(2)}[\xi]
+H^{(3)}[\xi] + \cdots$, where 
\begin{align}
  H^{(1)}[\xi] & = \int d^3x \, \left( -2 \partial^j \pi_{ij}\xi^i +
    ( \partial^i\partial^j h_{ij} - \Delta
    h) \xi^\perp \right)\\
  & = \int d^3x \, \left( \cH_m^{(1)}\xi^m+\cH^{(1)}_\perp
    \xi^\perp\right)
\end{align}
are the gauge generators associated to the constraints \eqref{eq:2a}
of the Pauli-Fierz theory. 
Because 
\begin{equation}
H[[\xi,\eta]_{SD}]=H^{(1)}[[\xi,\eta]_{SD}^{(0)}]+H^{(2)}[[\xi,\eta]_{SD}^{(0)}]
+H^{(1)}[[\xi,\eta]_{SD}^{(1)}]+O(z^3),\label{eq:exprhs}
\end{equation}
we have to lowest non trivial order
\begin{equation}
  \label{eq:50}
 \left\{H^{(1)}[\xi],H^{(2)}[\eta]\right\}=H^{(1)}[[\xi,\eta]_{SD}^{(0)}].
\end{equation}
This means that $H^{(2)}[\eta]$ are observables, i.e., weakly gauge
invariant functionals.

One can use integrations by parts to show that
$H^{(1)}[\xi_P]=0$. It then follows that 
\begin{equation}
\left\{ H[\xi_P] , H[\eta_P]\right\} = \left\{ H^{(2)} [\xi_P], 
H^{(2)}[\eta_P]\right\} + O(z^3).
\end{equation}
For vectors $\xi_P(\tilde\omega,\tilde a),\eta_P(\tilde \theta,\tilde
b)$ of the form \eqref{vect}, the first term on the RHS of
\eqref{eq:exprhs} vanishes on account of \eqref{vectalgebra}. To
lowest non trivial order, \eqref{Halgebra} then implies
\begin{equation}
  \label{eq:48}
  \left\{ H^{(2)} [\xi_P], 
H^{(2)}[\eta_P]\right\}=H^{(2)}[[\xi_P,\eta_P]_{SD}^{(0)}]
+H^{(1)}[[\xi_P,\eta_P]_{SD}^{(1)}]. 
\end{equation}
The generators $H^{(2)} [\xi_P]$ equipped with the Poisson bracket
thus form a representation of the Poincar\'e algebra when the
constraints of the Pauli-Fierz theory are satisfied. Explicitly, the
term proportional to the constraints is
\begin{equation}
  \label{eq:28}
  H^{(1)}[[\xi,\eta]_{SD}^{(1)}]=-2\int d^3x\, \d^j\pi_{ji} h^{ik}(
  \xi^\perp_P\theta^\perp_{\phantom{\perp}k} -\eta^\perp_P
\omega^\perp_{\phantom{\perp}k}),
\end{equation}
while
\begin{align}
\mathcal{H}^{(2)}_i & = - 
2 \partial_j \left(\pi^{jk} h_{ik} \right)+ \pi^{jk}\partial_i
h_{jk}\\
\mathcal{H}^{(2)}_\perp & =  \pi^{ij} \pi_{ij} - 
\frac{1}{2} \pi^2\nonumber\\
&+\frac{1}{4} \partial_k h_{ij} \partial^k h^{ij} -
\frac{1}{2} \partial_k h^{ki} \partial^j h_{ij} +
\frac{1}{2} \partial_i h \partial_j h^{ij} 
- \frac{1}{4} \partial_i h \partial^i h\nonumber\\
&+\partial_l\left( \frac{1}{2} h \partial^l h 
- h^{ij}\partial^l h_{ij} - \frac{1}{2} h \partial_i h^{il} 
- h^{il}\partial_i h + \frac{3}{2} h^{lj}\partial^i h_{ij}
+\frac{1}{2} h_{ij} \partial^i h^{jl}\right).
\end{align}

Isolating terms proportional to the constraints, we find
\begin{align}
  \label{eq:36}
  H^{(2)}[\xi]&=\int d^3x\, \left(\cH_m h^{mi}\xi_i+\frac{1}{2} \cH h
    \xi^\perp\right)  + \bar H^{(2)}[\xi],\\
\bar \cH^{(2)}_i&=-\pi^{jk}(\d_jh_{ki}+\d_k h_{ji}-\d_ih_{jk}),\\
\bar \cH^{(2)}_\perp&=\pi^{ij} \pi_{ij} - 
\frac{1}{2} \pi^2\nonumber\\
&+\frac{1}{4} \partial_k h_{ij} \partial^k h^{ij} -
\frac{1}{2} \partial_k h^{ki} \partial^j h_{ij} +  \frac{1}{4}\partial_i
h \partial^i h 
\nonumber\\
&+\partial_l\left(
- h^{ij}\partial^l h_{ij} 
- h^{il}\partial_i h + \frac{3}{2} h^{lj}\partial^i h_{ij}
+\frac{1}{2} h_{ij} \partial^i h^{jl}\right),
\end{align}
with $\bar H^{(2)}[\xi]=\int d^3x\,\left( \bar \cH^{(2)}_i\xi^i+\bar
  \cH^{(2)}_\perp \xi^\perp\right)$.  On account of \eqref{eq:50} and
the analog of \eqref{eq:32} for $H^{(1)}[f]$, it follows that
\begin{equation}
  \label{eq:51}
  \left\{ \bar H^{(2)} [\xi_P], 
\bar H^{(2)}[\eta_P]\right\}\approx \bar H^{(2)}[[\xi_P,\eta_P]_{SD}^{(0)}],
\end{equation}
where $\approx$ means an equality up to terms proportional to the
constraints $\cH_m,\cH_\perp$ of Pauli-Fierz theory. Note that the
functionals $H^{(2)}[\xi_P]$ and $\bar H^{(2)}[\xi_P]$ generate
transformations of the canonical variables that are equivalent because
they differ at most by a gauge transformation of the Pauli-Fierz
theory when restricted to the constraint surface.

The generators for global Poincar\'e transformations of Pauli-Fierz
theory can then be identified as 
\begin{eqnarray}
  \label{eq:100}
&&  Q_G(\omega,a)=\half\omega_{\mu\nu}J^{\mu\nu}_G-a_\mu P^\mu_G
=\bar H^{(2)}[\xi_P(\tilde \omega,\tilde
  a)]\nonumber\\
&& \tilde\omega_{\mu\nu}=\omega_{\mu\nu},\quad \tilde a_\perp=
a_\perp, \quad \tilde a_i= a_i+\omega_{\perp i}x^0.
\end{eqnarray}
Indeed, differentiating \eqref{eq:48} with respect to $b_\perp$ gives
\begin{eqnarray}
  \label{eq:103}
  \{H,Q_G(\omega,a)\}=\ddl{}{t}Q_G(\omega,a)+2\int d^3x\, \d^j\pi_{ji} h^{ik}
\omega_{\perp k}.
\end{eqnarray}
When combined with \eqref{eq:48} and \eqref{eq:100}, this shows that,
on the constraint surface, the generators $Q_G(\omega,a)$ are
conserved and satisfy the Poincar\'e algebra.

Finally, we can further simplify the explicit expression for $\bar
H^{(2)}[\xi_P]$ by using linearity of $\xi_P$ in $x^i$ and
integrations by parts to show that
\begin{align}
  \label{eq:52}
\int d^3x\,\bar
 \cH^{(2)}_\perp \xi^\perp_P=\int d^3x\, \Big[&\pi^{ij} \pi_{ij} - 
\frac{1}{2} \pi^2+\frac{1}{4} \partial_k h_{ij} \partial^k h^{ij} -
\frac{1}{2} \partial_k h^{ki} \partial^j h_{ij}   \nonumber\\
&+\frac{1}{4}\partial_i
h \partial^i h +\partial_l\big(
h \partial_i h^{il} +  h^{lj}\partial^i h_{ij}\big)\Big]\xi^\perp_P.
\end{align}

The expansion of the gauge transformations \eqref{eq:47},
\eqref{eq:47a} gives to first order:
\begin{gather}
  \label{eq:49}
  \delta^{(0)}_\xi h_{ij}= \d_i\xi_j+\d_j\xi_j,\qquad \delta^{(0)}_\xi
  \pi^{ij}=(\d^i\d^j-\delta^{ij} \Delta)\xi^\perp,\\
\delta^{(1)}_\xi h_{ij}=\xi^k\d_k h_{ij}+\d_i \xi^k h_{kj}+\d_j\xi^k
h_{ik} +2 \pi_{ij}\xi^\perp-\delta_{ij}\pi\xi^\perp,\label{eq:49a}\\
\delta^{(1)}_\xi \pi^{ij}=\half h(\d^i\d^j
-\delta^{ij}\Delta)\xi^\perp
-h^{im}\d_m\d^j\xi^\perp-h^{jm}\d_m\d^i\xi^\perp+h^{ij}
\Delta\xi^\perp\nonumber\\+\delta^{ij}h^{mn}
\d_m\d_n\xi^\perp+\d_m(\pi^{ij}\xi^m)-\pi^{mj}\d_m\xi^i
-\pi^{mi}\d_m\xi^j
\nonumber\\ +\half \d_k\xi^\perp\Big[
-\d^j h^{ki}-\d^i h^{kj} +\d^k h^{ij}+\delta^{ij}(2\d_l h^{kl}-\d^k
h)\Big]
\nonumber\\+\half \xi^\perp\Big[\d^i\d^j h
+\Delta h^{ij} -\d_k \d^i h^{jk} -\d_k \d^j
h^{ik} -\delta^{ij} (\Delta h-\d_k\d_l h^{kl})\Big]. \label{eq:49b}
\end{gather}

\section{Riemann tensor and canonical variables}
\label{sec:decomp-line-riem}

By following \cite{Henneaux:2004jw,Bunster:2006rt} (up to
conventions), we show in this appendix that the duality rotations that
we have defined coincide on-shell with the standard simultaneous
duality rotations among the (linearized) Riemann tensor and its dual
together with those for the electric and magnetic conserved
sources. We do this by showing how the covariant Riemann tensor is
expressed in terms of the canonical variables. This gives us the
appropriate generalization of the Gauss-Codazzi relations in the case
of both electric and magnetic sources.

\subsection{Covariant equations in the presence of magnetic sources}
\label{sec:covar-equat-pres}

Our conventions are as follows. Define $\epsilon_{a_1\dots
  a_n}=\epsilon^{a_1\dots a_n}$ to be totally skew-symmetric with
$\epsilon_{1\dots n}=1$. The Levi-Civita tensor is $\lc_{a_1\dots
  a_n}=\sqrt{|g|} \epsilon_{a_1\dots a_n}$. Indices on this tensor are
raised with the metric, which implies that $\lc^{a_1\dots
  a_n}=\frac{(-)^\sigma}{\sqrt{|g|}} \epsilon^{a_1\dots a_n}$ where
$\sigma$ is the signature of the metric. Our convention for the dual
is ${}^*\!\omega_{a_1\dots a_{n-p}}=\frac{1}{p!}  \omega^{b_1\dots
  b_p}\lc_{b_1\dots b_p a_1\dots a_{n-p}}$.

In flat Minkowski spacetime, traces are taken with the flat Minkowsi
metric $\eta_{\mu\nu}=\text{diag}(-1,1,1,1)$. Start with the
``(linearized) Riemann tensor'' $R^1_{\mu\nu\rho\sigma}\equiv
R_{\mu\nu\rho\sigma}$, with only symmetry properties skew-symmetry in
the first and last pairs of indices,
\begin{equation}
  \label{eq:53a}
  R_{\mu\nu\rho\sigma}  =  - R_{\nu\mu\rho\sigma} =
  -R_{\mu\nu\sigma\rho}.
\end{equation}
Its double dual 
(cf.~MTW \cite{Misner:1970aa}) 
\begin{equation}
\nG^1_{\mu\nu\rho\sigma}\equiv \nG_{\mu\nu\rho\sigma}=\frac{1}{4}
{\lc_{\mu\nu}}^{\alpha\beta}{R_{\alpha\beta}}^{\gamma\delta}
\lc_{\gamma\delta\rho\sigma}\label{eq:83},
\end{equation}
has the same symmetry properties. If ordinary duals of
$R_{\mu\nu\rho\sigma}$ and $\nG_{\mu\nu\rho\sigma}$ are taken with
respect to the last pair of indices, and we define
\begin{gather}
R^2_{\mu\nu\rho\sigma}\equiv-{}^*R_{\mu\nu\rho\sigma}=-\half
{R_{\mu\nu}}^{\alpha\beta}\lc_{\alpha\beta\rho\sigma},\\
\nG^2_{\mu\nu\rho\sigma}\equiv-{}^*\nG_{\mu\nu\rho\sigma}=\half
{\lc_{\mu\nu}}^{\alpha\beta}R_{\alpha\beta\rho\sigma},
\label{eq:76}
\end{gather}
we have
\begin{equation}
  \label{eq:70b}
  R^a_{\mu\nu\rho\sigma}=\epsilon^{ab}({}^*R)_{b\,\mu\nu\rho\sigma},\quad 
\nG^a_{\mu\nu\rho\sigma}=\epsilon^{ab}{}^*\nG_{b\,\mu\nu\rho\sigma}.
\end{equation}
The $36$ independent components of the Riemann tensor can be encoded
in 
\begin{equation}
  \label{eq:57}
  R^a_{0m0n},\quad \nG^a_{0m0n}. 
\end{equation}

The Ricci and Einstein tensors are defined as
\begin{equation}
R^a_{\mu\nu}={R^{a\alpha}}_{\mu\alpha\nu},\quad
G^a_{\mu\nu}={\nG^{a\alpha}}_{\mu\alpha\nu}=R^a_{\nu\mu}-\half
\eta_{\mu\nu}R^a\label{eq:77}. 
\end{equation}
If electric and magnetic conserved sources are
$T^{\mu\nu}_a\equiv(T^{\mu\nu},\Theta^{\mu\nu})$, with
$T^{\mu\nu}_a=T^{\nu\mu}_a$ symmetric, $\d_\mu T^{\mu\nu}_a=0$, the
duality rotations are defined by
\begin{gather}
  R^{a\prime}_{\mu\nu\rho\sigma}  =  {M^a}_b  R^b_{\mu\nu\rho\sigma}, \qquad
  T^{a\prime}_{\mu\nu}  = {M^a}_b  T^b_{\mu\nu} ,\nonumber \\
  M_{ca}{M^c}_b  =\delta_{ab}. 
\end{gather}
For a tensor $K^{\mu\nu}$, let
$\bar K^{\mu\nu}=K^{\mu\nu}-\half \eta^{\mu\nu} K$. 

It has been shown in \cite{Bunster:2006rt} that the duality invariant
equations of motion are
\begin{equation}
  \label{eq:75}
  G^{\mu\nu}_a=8\pi G\, T^{\mu\nu}_a\iff R^a_{\mu\nu\rho\sigma} + 
R^a_{\mu\sigma\nu\rho} +R^a_{\mu\rho\sigma\nu} = 8\pi G\, \epsilon^{ab}
  \lc_{\delta\nu\rho\sigma}\, 
  \overline{T}^\delta_{b\,\mu}. 
\end{equation}
They imply in particular that, on-shell, the tensors
$R^a_{\mu\nu\rho\sigma}$ are symmetric in the exchange of pairs of
indices and that $R^a_{\mu\nu}, G^a_{\mu\nu}$ are symmetric.
Furthermore, the Bianchi ``identities'' read
\begin{multline}
\partial_\epsilon R^a_{\gamma\delta\alpha\beta} +\partial_\beta 
R^a_{\gamma\delta\epsilon\alpha} +\partial_\alpha 
R^a_{\gamma\delta\beta\epsilon} =  8\pi G\, \epsilon^{ab}
\lc_{\epsilon\alpha\beta\rho} 
\left( \partial_\gamma \overline{T}^\rho_{b\,\delta} - 
\partial_\delta \overline{T}^{\rho}_{b\,\gamma}\right)\\ \iff
  \partial_\mu R_a^{\gamma\delta\rho\mu}  =  8\pi G\, \left(\partial^\delta
\overline{T}_a^{\rho\gamma}-
\partial^\gamma\overline{T}_a^{\rho\delta} \right),\label{eq:bianch}
\end{multline}
while the contracted Bianchi identities are
\begin{equation}
  \label{eq:58}
  \d_\nu G_a^{\mu\nu}=0. 
\end{equation}

Let 
\begin{multline}
  \label{eq:78}
  {K^{\lambda\tau}}_{\mu\nu\rho\sigma}
  [R^a_{\lambda\tau}]=\frac{1}{2} \left[ \eta_{\mu\rho} R^a_{\nu\sigma} +
    \eta_{\nu\sigma} R^a_{\mu\rho} - \eta_{\mu \sigma} R^a_{\nu\rho} -
    \eta_{\nu\rho} R^a_{\mu\sigma}\right] -\\- \frac{R^a}{6} \left[
    \eta_{\mu\rho} \eta_{\nu\sigma} -  \eta_{\mu \sigma}
    \eta_{\nu\rho}\right]. 
\end{multline}
Defining
\begin{equation}
  \label{eq:84}
  \tilde R^a_{\mu\nu\rho\sigma}=R^a_{\mu\nu\rho\sigma}-\half
\epsilon^{ab}\lc_{\rho\sigma\alpha\beta}
{{K^{\lambda\tau}}_{\mu\nu}}^{\alpha\beta}[R_{b\lambda\tau}],
\end{equation}
the tensor $\tilde R^a_{\mu\nu\rho\sigma}$ is skew in the first and
last pairs of indices, satisfies the cyclic identity because
$\lc^{\gamma\nu\rho\sigma}
R^a_{\mu\nu\rho\sigma}=\lc^{\gamma\nu\rho\sigma}\half
\epsilon^{ab}\lc_{\rho\sigma\alpha\beta}
{{K^{\lambda\tau}}_{\mu\nu}}^{\alpha\beta}[R_{b\lambda\tau}]$ and, as
a consequence, is also symmetric in the exchange of the first and last
pair of indices, $\tilde R^a_{\mu\nu\rho\sigma}= \tilde
R^a_{\rho\sigma\mu\nu}$. The associated Ricci tensors $\tilde
R^a_{\nu\sigma}=
R^a_{\nu\sigma}-\half\epsilon^{ab}\lc_{\nu\sigma\mu\alpha}R^{\mu\alpha}_b$
is then symmetric, $\tilde R^a_{\nu\sigma}= \tilde R^a_{\sigma\nu}$.
It follows that $\tilde R^a_{\nu\sigma}=R^a_{(\nu\sigma)}$ and
$R^a_{[\nu\sigma]}=\half\epsilon^{ab}\lc_{\nu\sigma\mu\alpha}R^{\mu\alpha}_b$.
The Weyl tensors are then defined as usual in terms of $\tilde
R^a_{\mu\nu\rho\sigma}$,
\begin{equation}
  \label{eq:59}
C^a_{\mu\nu\rho\sigma}=\tilde R^a_{\mu\nu\rho\sigma}-
{K^{\lambda\tau}}_{\mu\nu\rho\sigma}[\tilde R^a_{\lambda\tau}],
\end{equation}
and satisfy all standard symmetry properties: skew-symmetry in the first
and last pairs of indices, tracelessness (because
$\tilde R^a_{\nu\sigma}={K^{\lambda\tau\,
    \mu}}_{\nu\mu\sigma}[\tilde R^a_{\lambda\tau}]$), the cyclic identity
(because
$\epsilon^{\gamma\nu\rho\sigma}{K^{\lambda\tau}}_{\mu\nu\rho\sigma}
[\tilde R^a_{\lambda\tau}]=0$), which implies also symmetry in the
exchange of the first and last pair of indices,
\begin{gather}
  \label{eq:80}
  C^a_{\mu\nu\rho\sigma}  =  - C^a_{\nu\mu\rho\sigma} =
  -C^a_{\mu\nu\sigma\rho},\\
C^{\mu a}_{\nu\mu\sigma}=0,\quad
\epsilon^{\gamma\nu\rho\sigma}C^a_{\mu\nu\rho\sigma}=0, \quad 
C^a_{\mu\nu\rho\sigma}  =   C^a_{\rho\sigma\mu\nu}. 
\end{gather}
As usual, the 10 independent components of the Weyl tensor can be
parametrized by the electric and magnetic components
$E^a_{mn}\equiv(E_{mn},B_{mn})$, symmetric and traceless tensors
defined by 
\begin{equation}
  \label{eq:79}
  E^a_{mn}=C^a_{0m0n}=\half \epsilon_{njk}\epsilon^{ab}
  C_{b\,0m}^{\phantom{\mu\nu}jk}.
\end{equation}

Putting all definitions together, the relation between the Riemann and
Weyl tensors is
\begin{align}
  \label{eq:86}
R^a_{\mu\nu\rho\sigma}&=C^a_{\mu\nu\rho\sigma}+
{K^{\lambda\tau}}_{\mu\nu\rho\sigma}[R^a_{\lambda\tau}]+\half
\epsilon^{ab}\lc_{\rho\sigma\alpha\beta}
{{K^{\lambda\tau}}_{\mu\nu}}^{\alpha\beta}[R_{b(\lambda\tau)}]\\
&=C^a_{\mu\nu\rho\sigma}+
{K^{\lambda\tau}}_{\mu\nu\rho\sigma}[R^a_{(\lambda\tau)}]+\half
\epsilon^{ab}\lc_{\rho\sigma\alpha\beta}
{{K^{\lambda\tau}}_{\mu\nu}}^{\alpha\beta}[R_{b\lambda\tau}].
\end{align}
In particular, it follows that the $36$ independent components of the
Riemman tensor $R^1_{\mu\nu\rho\sigma}$ can be parameterized by the
$10$ independent components of the Weyl tensor
$C^1_{\mu\nu\rho\sigma}$, the $16$ components of the Ricci
tensor $R^1_{\lambda\tau}$, and the $10$ components of
  $R^2_{(\lambda\tau)}$. 

If we define
\begin{equation}
  \label{eq:87}
  \cE^a_{mn}=R^a_{0(m|0|n)}, \quad \cF^{am}=\half \epsilon^{mjk} 
R^a_{0[j|0|k]},\quad \cR^a_{mn}=R^a_{(mn)}+\cE^a_{mn}
\end{equation}
the parameterization consisting in choosing the symmetric tensors
$\cE^a_{mn},\cR^a_{mn}$ (24 components), $\cF^a_m$, (6 components),
and $R^1_{[\mu\nu]}(={}^*R^2_{[\mu\nu]})$ (6 components) is more
useful for our purpose.  That all tensors can be reconstructed from
these variables follows from the fact that
\begin{gather}
  \label{eq:88}
  R^a_{0m}=-2\epsilon^{ab}\cF_{bm},\quad
  R^a_{00}=\cE^{a},\quad R^a_{(mn)}=\cR^a_{mn}-\cE^a_{mn}. 
\end{gather}
This means that the symmetric part of the Ricci tensors can be
reconstructed from the variables. Since the antisymmetric parts belong
to the variables, so can the complete Ricci tensors $R^a_{\mu\nu}$.
Using now \eqref{eq:86} and definitions
\eqref{eq:79}, \eqref{eq:87}, \eqref{eq:78}, we find
\begin{equation}
  \label{eq:89}
  E^a_{mn}=\half(\cE^a_{mn}+\cR^a_{mn})-\frac{\delta_{mn}}{6}(\cE^{a}+ 
\cR^{a}).
\end{equation}
It follows that the Weyl tensors and then, using again \eqref{eq:86},
the Riemann tensors can  be reconstructed. 

In terms of the new parameterization, the equations of motion
\eqref{eq:75} read $R^a_{[\mu\nu]}=0$ and
\begin{gather}
  \label{eq:90}
  -2\epsilon^{ab}\cF_{bm}=8\pi G T^{a}_{0m}, \\
  \half \cR^{a}=8\pi G T^a_{00},\\
  \cR^a_{mn}-\cE^a_{mn}+\delta_{mn}(\cE^{a}-\half 
  \cR^{a})=8\pi G T^a_{mn}. \label{eq:90b}
\end{gather}
Using these equations of motion, the Bianchi identities
\eqref{eq:bianch} are equivalent to 
\begin{equation}
  \label{eq:53}
  \d^k(\epsilon_{ikm}\cF^{am}+\cR^a_{ik})=\half\d_i\cR^a,
\end{equation}
\begin{equation}
  \label{eq:91a}
  2\epsilon^{ab}\d_0\cF_{bm}=\d^n
  (\cE^a_{mn}+\epsilon_{mnk} \cF^{ak})-\d_m\cE^a, 
\end{equation}
\begin{multline}
  \label{eq:92}
  \d_0\cR^a_{ik}=\half\epsilon^{ab}\big[
   \epsilon_{kjl}\d^j{\cE_{b i}}^l+
\epsilon_{ijl}\d^j{\cE_{b k}}^l-2\delta_{ik}\d_j\cF^j_b-\d_i
\cF_{bk}-\d_k\cF_{bi}
\big]\iff\\
\epsilon^{ab}\d_0(\cR^{ik}_{b}-\half\delta^{ik}
\cR_b)=-\half
\big[\epsilon^{klm}\d_l{\cE^a_{m}}^i+\epsilon^{ilm}
\d_l{\cE^a_{m}}^k+2\delta^{ik}\d^j
\cF^a_j-\d^i\cF^{ak}-\d^k\cF^{ai}\big].
\end{multline}

\subsection{Canonical expressions}
\label{sec:riemann-tensor-terms}

We will now express the Riemann tensor in terms of the canonical
variables in such a way that the covariant equations
\eqref{eq:90}-\eqref{eq:92} coincide with the Hamiltonian equations
deriving from \eqref{eq:61}. 

From the constraints with sources, we find
\begin{gather}
  \label{eq:70}
  \cR^a=\d^m\d^n h^a_{mn}-\Delta h^a=-\Delta^2 C^a,\\
\cF^a_{m}=\half\Delta\d^n H^a_{mn}.  \label{eq:70a}
\end{gather}
Assuming $\Delta$ to be invertible, which we do in the rest of this
appendix, $\cR^a$ and $C^a$, respectively $\cF^a_m$ and $\d^n
H^a_{mn}$ determine each other. By taking the divergence, the Bianchi
identity \eqref{eq:53} implies that
\begin{equation*}
\d^m\d^n\cR^a_{mn}=-\half\Delta^3 C^a.
\end{equation*}
Similarily, the Bianchi identity \eqref{eq:91a} implies in particular
that $\Delta\cE^a-\d^m\d^n\cE^a_{mn}=\epsilon^{ab}\d_0\Delta\d^m\d^n
H_{bmn}$. When combined with \eqref{eq:90b}, the equations of motion
following from variation with respect to $C^a$ read
\begin{equation*}
  \frac{1}{2}\Delta^3 C^a+ \epsilon^{ab} \Delta\d_0
  (\Delta H_b-\d^m\d^n
  H_{bmn})+ 2\Delta^2 n^a=
\Delta\cE^a-\d^m\d^n(\cR^a_{mn}-\cE^a_{mn}).
\end{equation*}
When combined with the previous relations, they imply that 
\begin{gather*}
  \label{eq:85}
  \cE^a= -\half \epsilon^{ab}\d_0\Delta H_b+ \Delta n^a,\\
  \d^m\d^n\cE^a_{mn}= -\half\epsilon^{ab}\d_0\Delta(\Delta
  H_b-2\d^m\d^n H_{bmn})+ \Delta^2 n^a.
\end{gather*}
The rest of the Bianchi identities \eqref{eq:53}, \eqref{eq:91a} are
taken into account by applying a curl. This gives
$\epsilon^{rsi}\d_s \d^k\cR^a_{ik}=\half \Delta (\Delta\d^k
H^{ar}_k-\d^r \d^m\d^n H^a_{mn})$ and $\epsilon^{rsi}\d_s
\d^k\cE^a_{ik}=\epsilon^{rsi}2\epsilon^{ab}\d_0\d_s\cF_{bi}-
\d^r\d^k \cF^a_k+\Delta \cF^{ar}$. Yet another curl gives
$\d_k\d^m\d^n \cR^a_{mn}-\Delta \d^n\cR^a_{kn}=\half
\epsilon_{klr}\d^l \Delta^2 \d^n H^{ar}_n$ and $\d_k\d^m\d^n
\cE^a_{mn}-\Delta \d^n\cE^a_{kn}=2\epsilon^{ab}\d_0(\d_k \d^n \cF_{bn}
-\Delta \cF_{b k})+ \epsilon_{klr}\d^l \Delta
\cF^{ar}$. Using the previous relations we then get
\begin{gather*}
  \d^n\cR^a_{kn}=-\half \d_k \Delta^2 C^a-\half
  \epsilon_{klr}\d^l \Delta \d^n H^{ar}_n,\\
  \d^n\cE^a_{kn}= \epsilon^{ab}\d_0\Delta(-\frac{1}{2}\d_k
  H_b+\d^n H_{bkn})+ \d_k \Delta n^a-\half \epsilon_{klr}\d^l
  \Delta \d^n H^{ar}_n.
\end{gather*}

The equations of motion following from variation with respect to
$A^a_m$ are then identically satisfied. 

Defining $\cD^a_{mn}=\cR^a_{mn}-\cE^a_{mn}$ and using definition
\eqref{eq:3} of $\cP^{TT}$ combined with \eqref{eq:90b}, the equations
of motion following from variation with respect to $H^a_{mn}$ read
\begin{multline}
\epsilon_{ab}\d_0\Big[2 \left( \cP^{TT} H^b\right)_{mn} + \d_m \Delta
A_n^b +\d_n \Delta A_m^b +\half(\delta_{mn}\Delta-\d_m\d_n)C^b
\Big]-\epsilon_{ab} \Delta (\d_m n^b_n+\d_n n^b_m)-\\-2\Delta^2 H^a_{mn}
+\delta_{mn} \Delta^2 H^a
=-{\epsilon_{mpq}}\d^p\cD^q_{an}-{\epsilon_{npq}}\d^p\cD^q_{am}. \label{eq:eqH}
\end{multline}
Taking into account definition \eqref{eq:3} and previous relations, we
can extract
\begin{multline}
  \label{eq:81}
  -\Delta^{-1}\left(\cP^{TT} \cD^a\right)_{mn}=
\half \epsilon_{ab}\d_0\Big[2\left( \cP^{TT} H^b\right)_{mn}
-\epsilon_{mpq}\d_n \d^p \d^r H^{bq}_r-\epsilon_{npq}\d_m \d^p \d^r
H^{bq}_r +\\ + \d_m \Delta
A_n^b + \d_n \Delta A_m^b +\frac{1}{2}(\delta_{mn}\Delta-\d_m\d_n)C^b
\Big]-\epsilon_{ab} \Delta (\d_m n^b_n+\d_n n^b_m)-\\-\Delta^2 H^a_{mn}
+\half \delta_{mn} \Delta^2 H^a.
\end{multline}
In order to extract the remaining information from \eqref{eq:eqH}, 
we first apply $\delta^{mn}\Delta-\d^m\d^n$ to get 
\begin{equation}
  \label{eq:98}
  \epsilon_{ab}\d_0 \Delta C^b+2 \d^m\d^n H_{amn}=0,
\end{equation}
and then a divergence $\d^m$ giving
\begin{equation}
  \label{eq:114}
  \epsilon_{ab}\d_0 (\Delta A^b_n-\epsilon_{npq}\d^p\d^k H^{b
    q}_k)=\epsilon_{ab} \Delta n_n^b+2 \Delta \d^k H^k_{an}-\half \d_n
  \Delta H^a -\d_n \d^k\d^l H^a_{kl}. 
\end{equation}
We can now inject the latter relations into \eqref{eq:eqH} and use
\eqref{eq:6}, \eqref{eq:dec} to get
\begin{align}
  \label{eq:93}
  \cD_{mn}^{aTT}&=-\epsilon^{ab}  \d_0 \Delta
  H^{TT}_{bmn}-\left(\cP^{TT} H^a\right)_{mn}, \\
\cD_{mn}^{a}&=-\epsilon^{ab}\d_0\Delta 
\Big[ H_{bmn}-\half \delta_{mn} H_b\Big]  -\left(\cP^{TT}
  H^a\right)_{mn}-\d_m\d_n n^a-\nonumber \\ & \hspace{4cm}
-\frac{1}{4}(\delta_{mn}\Delta+\d_m\d_n)\Delta C^a.  \label{eq:93a}
\end{align}

Injecting into the second form of the last Bianchi identity
\eqref{eq:92} and using previous relations gives
\begin{multline}
  \label{eq:95}
  \epsilon_{ab}\d_0 \cR^b_{ij}=-\left( \cO \cR_a\right)_{ij}+\Delta^2 H^{TT}_{aij}
 +\frac{1}{4} \Delta \d_i \d^k H_{akj} +\frac{1}{4} \Delta \d_j \d^k
  H_{aki} -\half \d_i\d_j\d^k\d^l H_{akl}\\-\half
  \epsilon_{ab}\d_0\Big[\epsilon_{iqn} \d^q\Delta H^{bn}_j +
  \epsilon_{jqn} \d^q \Delta
  H^{bn}_i+\half(\delta_{ij}\Delta+\d_i\d_j)\Delta C^b
  \Big].
\end{multline}
Identifying the terms with time derivatives gives
\begin{gather}
  \label{eq:96}
  \cR^a_{ij}=-\half\Big[\epsilon_{iqn} \d^q\Delta
  H^{an}_j + \epsilon_{jqn} \d^q
  \Delta H^{an}_i+\half
  (\delta_{ij} \Delta+\d_i\d_j)\Delta C^a \Big]\nonumber\\
  =\half\Big[\d_i\d^k h^a_{kj}+\d_j\d^k h^a_{ki}-\d_i\d_j h^a-\Delta
  h^a_{ij}- \epsilon_{ikl}\d^k\d^p\d_j H^{al}_p- \epsilon_{jkl}
  \d^k\d^p\d_i H^{al}_p\Big].
\end{gather}
The terms without time derivatives in \eqref{eq:95} then cancel
identically.  Together with \eqref{eq:93a} this then finally gives
\begin{gather}
  \label{eq:97}
  \cE_a^{ij}=\epsilon_{ab}\d_0\Delta \Big[ H^{bij}-\half\delta^{ij}
  H^b\Big]+\d^i\d^j n_a- \half \epsilon^{ikl}\d_k\d^j
  \d^p H_{alp}-\half \epsilon^{jkl}\d_k\d^i \d^p H_{alp}\nonumber\\
  =-\epsilon_{ab}\d_0(\pi^{bij}-\half \delta^{ij}\pi^b)
+\d^i\d^j n_a- \half \epsilon^{ikl}\d_k\d^j \d^p H_{alp}
-\half \epsilon^{jkl}\d_k\d^i \d^p H_{alp}.
\end{gather}

\subsection{Riemann tensor for linearized Taub-NUT}
\label{sec:riem-tens-line}

Following for instance \cite{cohentannoudji:1989} section $A_1.2$ and
using a regularization in Fourier space, we find for the gravitational
dyon at rest at the origin discussed in
section~\bref{sec:point-part-grav},
\begin{align}
  \label{eq:99}
  \cR^a_{ij}&=GM^a\big[\frac{16\pi}{3}\delta_{ij}
  \delta^3(x)+\frac{\eta(r)}{r^3}(\delta_{ij}-\frac{3x_ix_j}{r^2})\big],\\
  \cE^a_{ij}&= GM^a\big[\frac{4\pi}{3}\delta_{ij}
  \delta^3(x)+\frac{\eta(r)}{r^3}(\delta_{ij}-\frac{3x_ix_j}{r^2})\big],
\end{align}
where $\eta(r)$ is a regularizing function that suppresses the
divergence at the origin and is $1$ away from the origin. We then find
\begin{gather}
  \label{eq:101}
  R^a_{00}=GM^a 4\pi \delta^3(x),\quad R^a_{ij}=GM^a 4\pi
  \delta_{ij} \delta^3(x),\\
E^a_{ij}=GM^a \frac{\eta(x)}{r^3}(\delta_{ij}-\frac{3x_ix_j}{r^2}),
\end{gather}
and all other components of $R^a_{\mu\nu}$ vanishing. For the Riemann
tensor, this implies
\begin{gather}
  \label{eq:102}
  R^a_{0i0j}=GM^a\big[\frac{4\pi}{3}\delta_{ij}
  \delta^3(x)+\frac{\eta(x)}{r^3}(\delta_{ij}-\frac{3x_ix_j}{r^2})\big],\\
R^a_{0ijk}=-\epsilon^{ab}{\epsilon_{jk}}^lGM^a\big[\frac{4\pi}{3}\delta_{il}
  \delta^3(x)+\frac{\eta(x)}{r^3}(\delta_{il}-\frac{3x_ix_l}{r^2})\big],
\end{gather}
with all other components obtained through the on-shell symmetries of
the Riemann tensor. This is the usual Riemann tensor for the
linearized Taub-NUT solution.  

\newpage


\def\cprime{$'$}
\providecommand{\href}[2]{#2}\begingroup\raggedright\endgroup

\end{document}